\documentclass[twocolumn,showpacs,preprintnumbers,amsmath,amssymb,superscriptaddress]{revtex4}

\usepackage{amsmath}
\usepackage{graphicx}
\usepackage{amsfonts}
\usepackage{amssymb}
\usepackage{epsfig}
\usepackage{psfrag}

\newcommand{\be}{\begin{equation}}
\newcommand{\ee}{\end{equation}}
\newcommand{\beq}{\begin{eqnarray}}
\newcommand{\eeq}{\end{eqnarray}}

\def\nue{\mathrel{{\nu_e}}}
\def\numu{\mathrel{{\nu_\mu}}}
\def\nutau{\mathrel{{\nu_\tau}}}

\def\barnue{\mathrel{{\bar \nu}_e}}
\def\barnumu{\mathrel{{\bar \nu}_\mu}}

\def \lta {\mathrel{\vcenter{\hbox{$<$}\nointerlineskip\hbox{$\sim$}}}}
\def \gta {\mathrel{\vcenter{\hbox{$>$}\nointerlineskip\hbox{$\sim$}}}}

\def\t13{\mathrel{{\theta_{13}}}}
\def\y12{\mathrel{{\tan^2 \theta_{12}}}}
\def\c2{\mathrel{{\chi^2 }}}

\def\epee{\mathrel{{\epsilon_{ee}}}}
\def\epem{\mathrel{{\epsilon_{e\mu}}}}
\def\epmm{\mathrel{{\epsilon_{\mu\mu}}}}
\def\epet{\mathrel{{\epsilon_{e\tau}}}}
\def\epmt{\mathrel{{\epsilon_{\mu\tau}}}}
\def\eptt{\mathrel{{\epsilon_{\tau\tau}}}}

\begin{document}

\title{A test of tau neutrino interactions with atmospheric neutrinos and K2K}

\author{Alexander Friedland} \email{friedland@lanl.gov}
\affiliation{Theoretical Division, T-8,
  MS B285, Los Alamos National Laboratory, Los Alamos, NM 87545}

\author{Cecilia Lunardini} \email{lunardi@phys.washington.edu}
\affiliation{Institute for Nuclear Theory and University of
  Washington, Seattle, WA 98195}

\begin{abstract}
  The presence of a tau component in the flux of atmospheric neutrinos
  inside the Earth, due to flavor oscillations, makes these neutrinos
  a valuable probe of interactions of the tau neutrino with matter.
  We study -- analytically and numerically -- the effects of
  nonstandard interactions in the $\nu_e-\nu_\tau$ sector on
  atmospheric neutrino oscillations, and calculate the bounds on the
  exotic couplings that follow from combining the atmospheric neutrino
  and K2K data. We find very good agreement between numerical results
  and analytical predictions derived from the underlying
  oscillation physics. While improving on existing accelerator
  bounds, our bounds still allow couplings of the size comparable to
  the standard weak interaction.  The inclusion of new interactions
  expands the allowed region of the vacuum oscillation parameters
  towards smaller mixing angles, $0.2 \lta \sin^2 \theta_{23}\lta
  0.7$, and slightly larger mass squared splitting, $1.5 \cdot 10^{-3}
  ~{\rm eV^2}\lta |\Delta m^2_{23}| \lta 4.0 \cdot 10^{-3} ~{\rm
    eV^2}$, compared to the standard case.  The impact of the K2K data
  on all these results is significant; further important tests of
  the $\nu_e-\nu_\tau$ exotic couplings will come from neutrino beams
  experiments such as MINOS and long baseline projects.
\end{abstract}

\preprint{LA-UR-05-3140}
\maketitle

%
\section{Introduction}
\label{intro}

Despite the remarkable successes of the Standard Model (SM), it is
widely believed that there is new physics at the TeV scale, which
stabilizes the Higgs mass against large radiative corrections.  The
search for this physics has been the goal of many key particle physics
experiments of the last two decades. This search has been, and is
being, carried out in two, often complementary, directions: (i)
efforts involving colliding particles at highest achievable
energies, and (ii) efforts involving precision measurements at low
energies.

Here, we concentrate on the second possibility.  There are many
well-known examples of the low-energy techniques being very effective. For
instance, precision measurements of atomic parity violation have
resulted in an accurate determination of the interactions between the
electron and $u$ and $d$ quarks mediated by the $Z$-boson
\cite{AtomicPhysRep}. Searches for exotic decays of the muon
\cite{mutoegamma} and precision measurements of its anomalous dipole
moment \cite{gminus2} are placing important constraints on new physics
at the TeV scale. Finally, searches for proton decay \cite{SKproton}
are sensitive to certain types of new physics all the way up to the
scale of Grand Unification.

This paper deals with another probe of this type, namely, the process
of neutrino refraction in matter. More specifically, we will consider
the refraction of the atmospheric neutrinos in the matter of the Earth
and investigate the sensitivity of this process to new physics above
the electroweak scale \footnote{The refraction process is indeed a
  low-energy one, not because of the energies of the particles
  themselves -- which reach hundreds of GeVs for atmospheric neutrinos
  -- but because refraction involves the neutrino-matter
  forward-scattering amplitude.}. The focus on neutrinos is motivated
by the fact that they are the least tested particles of the Standard
Model. While the charged lepton sector has been extensively probed for
many exotic modes (such as $\mu \rightarrow e + \gamma$, which has an
upper limit of $1.2 \times 10^{-11}$ \cite{mutoegamma} on its
branching ratio), our knowledge about the neutrino sector is not
nearly at the same level. While in certain classes of models it is
possible to relate the properties of the two sectors, such relations
do not have general character \cite{BerezhianiRossi}, making it
necessary to probe the interactions of neutrinos directly.

The most direct limits on the couplings of neutrinos with matter are
provided by experiments of scattering of neutrino beams on a target.
Scattering tests the neutrino-matter cross section; results of this
type are given by CHARM \cite{Vilain:1994qy} and NuTeV
\cite{Zeller:2001hh} and constrain mainly the coupling of the muon
neutrino to matter, as described in Sec. \ref{hamintro}. The
interactions of the tau neutrino, however, and of the electron
neutrino in some channels are still poorly restricted, being allowed
at the same order as the Standard Model ones
\cite{BerezhianiRossi,Davidson:2003ha}.  To this end, oscillation
experiments may possess an important advantage: while it is hard to
produce a beam of tau neutrinos in a laboratory, a very abundant flux
of tau neutrinos and/or antineutrinos is produced by oscillation from
muon or electron-flavored fluxes.  It is also important that
scattering results are subject to obvious degeneracies, for example in
the complex phases of flavor-changing couplings, since they measure
probabilities rather than amplitudes. Neutrino oscillations probe
different combinations of NSI parameters with respect to accelerators
and thus can help to resolve the degeneracies.

The realization that oscillation experiments are sensitive to
non-standard interactions (NSI) of the neutrino is, or course, very
old. In fact, already in his seminal work \cite{Wolfenstein:1977ue},
which laid the foundations for the MSW effect
\cite{Wolfenstein:1977ue,Mikheev:1986wj,Mikheev:1986gs}, Lincoln
Wolfenstein focused on the possibility that non-standard neutrino
interactions may change the flavor composition of the solar neutrino
flux. The idea was further developed in \cite{Valle1987,Guzzo:1991hi}
and many other subsequent works.  Recently, the idea to use both solar
and atmospheric neutrino oscillation as a way to \emph{measure} the
neutrino-matter interactions has been receiving progressively more
attention
\cite{Fornengo:2001pm,Guzzo:2001mi,Friedland:2004pp,Guzzo:2004ue,Gonzalez-Garcia:2004wg,Miranda:2004nb,deGouvea:2004va,Friedland:2004ah}.
The underlying physical argument is the following: both solar and
atmospheric neutrino fluxes have been measured over a range of
energies and are fit very well by neutrino oscillations. One may hope,
then, that the system is already overconstrained and the introduction
of new physics would break the fit, thus opening the possibility to
constrain physics beyond the Standard Model strongly.

The case of atmospheric neutrinos looks especially promising in this
respect. The Super-Kamiokande (SK) experiment has collected data on
the neutrino survival probability as a function of the zenith angle
over five decades in neutrino energy, $E_\nu\sim[0.1, 10^{4}]$ GeV.
All these data are very well fit with neutrino oscillations, with only
two parameters \cite{Fukuda:1998mi,Fukuda:2000np,Ashie:2004mr}.
Furthermore, this result is confirmed by an independent measurement of
the neutrinos from an accelerator beam at the K2K experiment
\cite{Aliu:2004sq}. Thus, it is very natural to use atmospheric
neutrinos as probes of NSI.

The first investigations of this type
\cite{Fornengo:2001pm,Guzzo:2001mi,Gonzalez-Garcia:2004wg} indeed gave
very strong bounds. It was shown that if the analysis is restricted to
two flavors \cite{Fornengo:2001pm,Gonzalez-Garcia:2004wg}, $\nu_ \mu$
and $\nu_\tau$, one can constrain NSI down to the level of a few
percent of the standard weak interaction couplings. It was shown
later, however, that it is essential that the analysis be done with
all three flavors \cite{Friedland:2004ah}: the physical arguments for
reducing the problem down to two flavors used in the case of the
standard oscillation analysis no longer apply, in general, when one
introduces NSI on top of vacuum oscillations.

Within the framework of a three-flavor analysis it was shown
\cite{Friedland:2004ah} that the bounds derived in the two-flavor
regime are relaxed when the $\nu_e$ generation is included. The
details of the argument are as follows. The high-energy sample is the
most sensitive to matter effects. This happens since the vacuum
oscillation Hamiltonian is inversely proportional to the neutrino
energy, while the matter contribution to the Hamiltonian is energy
independent. It turns out, however, that in a certain region of the
parameter space the oscillations of high-energy atmospheric neutrinos
regain the character of vacuum oscillations. The low-energy neutrinos
remain in the vacuum oscillation regime and an overall satisfactory
fit to the data can be achieved. For very large NSI, the fit is
eventually broken because the values of the oscillation parameters
preferred by the high- and low-energy parts of the data become
incompatible with each other.

Our analysis in \cite{Friedland:2004ah} was essentially limited to
demonstrating the above point. In this work, we present the first
comprehensive study of the effects of NSI in the $e -\tau$ sector on
the oscillations of atmospheric neutrinos. Our analysis here has
both a numerical and an analytical part. We scan the full three
dimensional space of the (effective) NSI couplings
$\epee,\epet,\eptt$, and present the allowed region (marginalized
over the vacuum oscillations parameters) in the space of these
quantities. We also discuss several generalizations, including
subdominant effects like those of the non-zero $\theta_{13}$ mixing
angle and of the smaller, ``solar'' mass splitting. A third
important aspect is that our analysis updates the previous ones by
including the most recent results from the K2K experiment
\cite{Aliu:2004sq}.

The text contains generalities in Sect. \ref{gen}, where we give a
general review of atmospheric neutrinos and neutrino oscillations with
NSI. In Sect.~\ref{sect:analyt} we treat the problem of atmospheric
neutrino oscillations in the presence of NSI. We describe various
reductions to two-flavor oscillations (Sect.  \ref{2nured}) and show
how these help to understand the physics of the sensitivity of
atmospheric neutrinos to NSI (Sect.~\ref{dom}).  In Sect.
\ref{sect:num} we present a detailed numerical analysis of the
problem, including the discussions of dominant (Sect.~\ref{mainres})
and subdominant effects (Sect.~\ref{sub}).  Our summary and
conclusions follow in Sect.  \ref{concl}.

\section{Generalities}
\label{gen}

\subsection{Neutrino oscillations, masses and mixings}
\label{minidef}

The results of nearly all \footnote{A possible exception is the LSND
result \cite{Athanassopoulos:1995iw,Athanassopoulos:1996jb},
currently being tested at MiniBOONE \cite{Ray:2004hp}.} available
neutrino experiments can be explained assuming that the three
known neutrinos, $\nue, \numu, \nutau$, undergo flavor
oscillations. As is well known,  this means that neutrinos have
flavor mixing and non-zero masses.

Let us denote by $\nu_i$ ($i=1,3$) the neutrino mass eigenstates,
and by $m_i$ their masses.  In vacuum, the time evolution of these
states is described by the kinetic Hamiltonian, which is given by
the relativistic dispersion relation: $E=\sqrt{p^2 + m^2}$.  The
phenomenon of oscillations depends on the difference of the
quantum phases of the mass eigenstates;  therefore, for the
purpose of describing oscillations one can neglect the overall
constant in the Hamiltonian. The Hamiltonian in the mass basis can
be then written as follows:
\begin{equation}
    \label{eq:vacHmass}
    H_{\rm vac}^{\rm diag}={\rm Diag} \left(
  -\Delta_\odot - \Delta , \Delta_\odot -\Delta , \Delta_\odot+\Delta \right)~.
\end{equation}
We use the notation $\Delta \equiv\Delta m^2_{32}/(4 E)$,
$\Delta_\odot \equiv\Delta m^2_{21}/(4 E)$, with $E$ being the
neutrino energy and $\Delta m^2_{ij} \equiv m^2_i - m^2_j$. Here
$m_i\ll E$ has been assumed.

The connection to the flavor basis $\nu_\alpha$ is provided by the
matrix $U$, defined as $\nu_\alpha = U_{\alpha i} \nu_i$.  We adopt
the standard parameterization for $U$ (see e.g.
\cite{Krastev:1988yu}), which contains a Dirac phase, $\delta$, and
the three mixing angles $\theta_{ij}$:
\begin{widetext}
\begin{equation}
  \label{eq:Umatr}
  U =
  \begin{pmatrix}
    1 & 0 & 0 \\
    0 & \cos \theta_{23} & \sin \theta_{23} \\
    0 & -\sin \theta_{23} & \cos \theta_{23} \\
  \end{pmatrix}
  \begin{pmatrix}
    \cos \theta_{13} & 0 & \sin \theta_{13} e^{-i \delta} \\
    0 & 1 & 0 \\
    -\sin \theta_{13} e^{i \delta} & 0 & \cos \theta_{13} \\
  \end{pmatrix}
  \begin{pmatrix}
    \cos \theta_{12} & \sin \theta_{12} & 0 \\
    -\sin \theta_{12} & \cos \theta_{12} & 0 \\
    0 & 0 & 1 \\
  \end{pmatrix} ~.
\end{equation}
\end{widetext}
While the mass squared splittings, $\Delta m^2_{ij}$, determine the
frequency of the oscillations, the mixing angles control their
amplitudes.  In vacuum, no oscillations happen if the matrix $U$
equals the identity (zero mixing angles) or if the three masses $m_i$
are equal (zero mass splittings).

In many physical situations, and in the assumption of purely standard
interactions, observations happen to depend mainly on one mixing and
one mass square splitting, while the other parameters give small
corrections.  Conventionally, $\theta_{12}$ and $\Delta
m^2_{21}$ are assigned to describe the oscillations of solar
neutrinos, while $\theta_{23}$ and $\Delta m^2_{23}$ are used for
atmospheric neutrinos.  The third angle, $\theta_{13}$, gives small
effects on both solar and atmospheric neutrinos; it is hoped to be
tested with future precision experiments involving neutrino beams. The
sign of $\Delta m^2_{23}$ distinguishes between the two physically
different configurations of the mass spectrum: the ``normal'' neutrino
mass hierarchy ($\Delta m^2_{23}>0$) and the ``inverted'' one ($\Delta
m^2_{23}<0$).

It should be stressed that the possibility of reducing both solar and
atmospheric neutrino evolution to two-state oscillations is an
absolutely non-trivial result and should not be taken for granted. It
relies on the measured smallness of the $\theta_{13}$ mixing angle,
and of $\Delta m^2_{21}$ relative to $\Delta m^2_{23}$. Moreover, it
crucially depends on the matter interactions being standard. In the
presence of nonstandard interactions, while the solar neutrino
analysis can still be done with two states \cite{Friedland:2004pp},
the atmospheric neutrino case \emph{requires} a full three-neutrino
treatment \cite{Friedland:2004ah}.

\subsection{Fluxes of atmospheric neutrinos and experimental results}
\label{minirev}

Let us summarize the essential features of the fluxes of atmospheric
neutrinos, and of the available data on them.  We also mention tests
of oscillations with other neutrino sources that are relevant to our
discussion.  The reader is referred to the literature for a more
complete review
\cite{Gaisser:1990vg,Gonzalez-Garcia:2002dz,Fukugita:2003en,Kajita:2004ga}.

Atmospheric neutrinos are products of the absorption of cosmic rays in
the atmosphere of the Earth.  They proceed from pion and kaon decays
and, thus, are produced in the muon and electron species.  Complex
numerical models (see e.g.
\cite{Gaisser:1996dz,Bugaev:1998bi,Fiorentini:2001wa,Battistoni:2002ew,Barr:2004br,Honda:2004yz})
have been developed to predict how the flux of these neutrinos
develops in the atmosphere; here we summarize the main features of
these models.

Both neutrinos and antineutrinos are produced in similar abundances.
The energy spectrum of neutrinos detected at Super-Kamiokande
spans several orders of magnitudes, from $\sim 0.1$ GeV to over a TeV.
In the absence of neutrino oscillations, for energies higher than
$\sim 1$ GeV, the muon (electron) neutrino flux is predicted to
decrease as $\sim E^{-3}$ ($\sim E^{-3.5}$) (see e.g.
\cite{Honda:2004yz}). At lower energy the spectrum is made more
complicated by several factors, such as geomagnetic effects and solar
modulations.

At energies $E \lta 1$ GeV most muons decay in flight before reaching
the ground. Correspondingly, for these energies, one expects at ground
level a ratio of fluxes in the muon and electron flavors of about 2.
This ratio increases with neutrino energy, as more and more muons
reach the ground without decaying. Numerical models give the
muon-to-electron neutrino ratio with a $\sim 5\%$ accuracy. By
comparison, the uncertainty on the individual fluxes is significantly
larger, $\sim 20\%$.

After a long and extensive effort of many different collaborations,
the Super-Kamiokande experiment has conclusively demonstrated that
atmospheric muon neutrinos undergo oscillations
\cite{Fukuda:1998mi,Fukuda:2000np,Ashie:2004mr}. The indication of
oscillations comes from the observation of an energy- and
zenith-dependent muon neutrino and antineutrino fluxes.  The different
muon data samples at SK, which refer to different energy windows, show
that at low energy (sub-GeV events) and intermediate energy (multi-GeV
events) the muon (anti)neutrino flux is suppressed at large zenith
angles, suggesting a distance-dependent effect.  The
distance-dependent suppression is best seen in the multi-GeV events,
due to the better alignment of the direction of the detected lepton
with that of the incoming neutrino.  In the highest energy sample
(through going muons, $E \sim 10^{1}-10^{4}$ GeV) the suppression is
reduced in size for all zenith angles. It is important to notice that
the electron neutrino flux is not tested in the same interval of
energy: the e-like event sample reaches at most $E \sim 10$ GeV,
beyond which absorption in the rock prevents detection.

Detailed analyses of the SK data strongly favor oscillations of
$\numu$ into $\nutau$, and, for purely standard interactions, give the
parameters
\begin{eqnarray}
|\Delta m^2_{32}|&\sim& (1.7 - 3.6) \cdot 10^{-3}~{\rm eV^2},\nonumber\\
\sin^2 2\theta_{23} &\sim&  0.85 -1.
\label{atmpar}
\end{eqnarray}
These numbers are taken from the 99\% confidence level (C.L.) contours
of the recent Super-Kamiokande paper, ref. \cite{Ashie:2005ik}; they
agree with the results of our analysis (see
Fig.~\ref{finalanswer} later in Sect.~\ref{mainres}).

Several alternative neutrino conversion scenarios have been ruled out.
For example, conversion into a purely sterile neutrino ($\nu_s$) has
been excluded, mainly due to the non-observations of the $\numu-\nu_s$
matter effects inside the Earth \cite{Fukuda:2000np}.
A scenario in which neutrinos have no masses but oscillate in the
Earth due to NSI is strongly disfavored too
\cite{Lipari:1999vh,Fogli:1999fs,Fornengo:2001pm}.  Non-oscillations
mechanisms, like decoherence or neutrino decay are in strong tension
with the data, especially with the presence of a minimum in the $L/E$
event distribution at SK \cite{Ashie:2004mr} (here $L$ is the distance
traveled by the neutrinos from production to detection).

Given the picture of $\numu\rightarrow\nutau$ oscillations, one may
expect $\nu_\tau$-induced events in the SK detector and, in fact,
there is a hint of the presence of such events \cite{Saji:2003fw}.

Additional support for atmospheric neutrino oscillations comes from
the MACRO \cite{Ambrosio:1998wu,Ambrosio:2000ja,Ambrosio:2001je} and
Soudan2 \cite{Allison:1996yb,Sanchez:2003rb} atmospheric neutrino
experiments, and -- very importantly -- from the K2K neutrino beam
experiment \cite{Ahn:2002up,Aliu:2004sq}. K2K measures the
flux and energy distribution (centered at $E \sim 1$ GeV ) of muon
neutrinos produced by an accelerator at a distance of 250 km from the
detector. The results evidence an oscillatory disappearance of
$\numu$, with a region of oscillation parameters compatible with
atmospheric results:
\begin{eqnarray}
|\Delta m^2_{32}|&\sim& (1.2 -5) \cdot 10^{-3}~{\rm eV^2},\nonumber\\
\sin^2 2\theta_{23} &\sim&  0.25 -1,
\label{k2kpar}
\end{eqnarray}  
(99\% C.L. interval). Small regions with larger mass
squared splitting ($few \cdot 10^{-2}~{\rm eV^2}$) are also
allowed by K2K at 99\% C.L.

In addition to $\theta_{23}$ and $\Delta m^2_{32}$, the other vacuum
parameters are relevant to atmospheric neutrinos, as they contribute
to subdominant effects.  For this reason, we briefly summarize the
status of the tests of these quantities.

The parameters $\theta_{12}$ and $\Delta m^2_{21}$ are measured with
solar neutrinos and the KamLAND experiment
\cite{Eguchi:2002dm,Araki:2004mb}.  A combined analysis of their data,
with standard interactions only, gives \cite{Araki:2004mb,SNO2005}:
\begin{eqnarray}
\Delta m^2_{21}&\simeq& (7 -  10) \cdot 10^{-5}~{\rm eV^2}, \nonumber\\
\tan^2
\theta_{12}&\simeq& 0.4 - 0.55,
\label{kamdmq}
\end{eqnarray}
which corresponds to the Large Mixing Angle (LMA) solution of
the solar neutrino problem.
Interestingly, in presence of NSI other regions of the
$\theta_{12}$-$\Delta m^2_{21}$ plane become allowed
\cite{Friedland:2004pp,Guzzo:2004ue,Miranda:2004nb}.

The mixing $\theta_{13}$ is strongly constrained by the
non-observation results from short base-line reactor experiments. The
most conservative limit compatible with the allowed interval of
$|\Delta m^2_{32}|$ is: 
\begin{equation} 
  \sin^2 \theta_{13} \lta 0.02~,
  \label{th13}
\end{equation} 
as given by the CHOOZ experiment at 90\% C.L.
\cite{Apollonio:1999ae,Apollonio:2002gd}, and supported by the results
of Palo Verde \cite{Boehm:2001ik}.

%

\subsection{NSI and the three neutrino oscillation Hamiltonian in matter}
\label{hamintro}

When neutrinos propagate in a medium, an interaction term has to be
added to the Hamiltonian to account for neutrino refraction in matter.
For this, we consider the Standard Model (SM) weak interactions, and
possible NSI, both flavor changing (FC) and flavor preserving (FP).
We can write the NSI Lagrangian in the form of effective four-fermion
terms:
\begin{widetext}
\begin{eqnarray}
L^{NSI} &=& - 2\sqrt{2}G_F (\bar{\nu}_\alpha\gamma_\rho\nu_\beta)
(\epsilon_{\alpha\beta}^{f\tilde{f} L}\bar{f}_L \gamma^\rho
\tilde{f}_L + \epsilon_{\alpha\beta}^{f\tilde{f}
R}\bar{f}_R\gamma^\rho \tilde{f}_{R})
+ h.c.~,  
\label{eq:lagNSI}
\end{eqnarray}
\end{widetext}
where $\epsilon_{\alpha\beta}^{f\tilde{f} L}$
($\epsilon_{\alpha\beta}^{f\tilde{f} R}$) denotes the strength of the
NSI between the neutrinos $\nu$ of flavors $\alpha$ and $\beta$ and
the left-handed (right-handed) components of the fermions $f$ and
$\tilde{f}$.

The operators in Eq.~(\ref{eq:lagNSI}) may arise from physics at a
high energy scale, with new, heavy scalars and gauge bosons. While a
theorist may prefer to see a concrete model leading to neutrino NSI,
from the experimental point of view it is impractical to test any
given model in isolation. The advantage of using effective low-energy
operators is that they encompass the experimental effects of a variety
of models, including, very importantly, those that have not been
thought of yet.  The four-fermion form is appropriate since propagator
corrections are negligible even at the highest energies of the
atmospheric neutrino spectrum, just like for the SM interactions.

The effect of standard and non-standard interactions on
neutrino propagation is given by the sum over the contributions of the
individual scatterers. This results in the Hamiltonian:
\begin{equation}
H_{\rm mat}=\sqrt{2}G_F n_e\begin{pmatrix}
1+\epee & \epsilon_{e\mu}^\ast & \epsilon_{e\tau}^\ast \\
\epem & \epmm& \epsilon_{\mu\tau}^\ast \\
\epet & \epmt & \eptt \\
\end{pmatrix},
\label{eq:ham}
\end{equation}
in the flavor basis, up to an irrelevant identity term.  Here $n_e$ is
the electron number density of the medium and the definition
$\epsilon_{\alpha\beta}\equiv
\sum_{f=u,d,e}\epsilon_{\alpha\beta}^{f}n_f/n_e$ accounts for above
mentioned sum.  We use
$\epsilon_{\alpha\beta}^{f}\equiv\epsilon_{\alpha\beta}^{fL}+\epsilon_{\alpha\beta}^{fR}$
and $\epsilon_{\alpha\beta}^{fP}\equiv\epsilon_{\alpha\beta}^{ffP}$,
because matter effects are sensitive only to the interactions that
preserve the flavor of the background fermion $f$ (required by
coherence \cite{Friedland:2003dv}) and, furthermore, only to the
vector part of that interaction.

As mentioned in the introduction, the $ee$, $e\tau$ and $\tau \tau$
NSI are the least constrained by direct measurements on neutrinos.  Of
these, we quote some results (valid at $90\%$ C.L.) from
\cite{Davidson:2003ha}, where each NSI coupling was analyzed isolated
from the others:
\beq
|\epsilon^{u L}_{\tau \tau}| \lta 1.4 \nonumber \\
-0.6 \lta \epsilon^{d R}_{ee} \lta  0.5 \nonumber \\
|\epsilon^{d L}_{e \tau}| \lta 0.5~.
\label{epbounds}
\eeq
We also have $|\epsilon^{e L}_{\mu \tau} | \lta 0.1 $ and
$|\epsilon^{e R}_{\mu \tau}| \lta 0.1 $, while all the other epsilons
have stronger bounds: $|\epsilon^{d L}_{\mu e} | \lta 8 \cdot
10^{-4}$, $|\epsilon^{e R}_{\mu \mu} | \lta 0.03$
\cite{Davidson:2003ha}.

We note that these bounds are much weaker than the corresponding ones
in the charged lepton sector. In many models, the latter can be
carried over to the neutrinos by the $SU(2)$ symmetry.  Since the
$SU(2)$ symmetry is violated, however, it is also possible that the
NSI couplings receive $SU(2)$-violating contributions.  For example,
it was shown that certain dimension eight operators involving the
Higgs field \cite{BerezhianiRossi,Davidson:2003ha} affect the
neutrinos but not the charged leptons. Thus, it is important to seek
direct, model-independent limits on the neutrino interactions, and our
work, it is hoped, contributes in this direction.

Motivated by the loose limits in Eq.~(\ref{epbounds}), we focus on NSI
in the $\nu_e-\nu_\tau$ sector, take $\epem=\epmt=\epmm=0$, and study
in detail the oscillations of atmospheric neutrinos with the NSI
described by $\epee,\epet,\eptt$.  Setting $\epem$ and $\epmm$ to zero
is certainly justified in view of the strong direct limits on these
couplings \cite{Davidson:2003ha}.  The case for $\epmt$ is a bit more
subtle.  Arguments can be made that even for $\epet\ne 0$ the
sensitivity of the data to $\epmt$ is essentially the same as in the
two neutrino analysis \cite{Fornengo:2001pm}, and therefore the bound
$|\epmt| \lta {\cal O}(10^{-2})$, found in \cite{Fornengo:2001pm},
should apply. We hope to return to this point in a future work.

\section{Atmospheric neutrinos and NSI: analytical treatment}
\label{sect:analyt}

\subsection{Reduction to two neutrinos}
\label{2nured}

As follows from Sect. \ref{minidef} and \ref{hamintro}, the
oscillations of neutrinos in matter are described, in the flavor
basis, by the Hamiltonian
\begin{equation}
H=U H_{\rm vac}^{\rm diag} U^\dagger +H_{mat}.
\end{equation}
In general, the density $n_e$ varies along the neutrino trajectory and
the resulting time dependence of the Hamiltonian is too complicated to
allow an exact solution of the Schroedinger equation. Moreover, in
presence of NSI on quarks there is a further dependence on the
nucleon-to-electron ratio, and therefore on the chemical composition
of the medium. In spite of this, for purely standard interactions the
oscillations of atmospheric neutrinos in the Earth is well
approximated by a two neutrino oscillation, $\numu \leftrightarrow
\nutau$ , on most of the energy spectrum.  This is thanks to the
smallness of $\theta_{13}$ and to the hierarchy of the mass splittings
( $| \Delta m^2_{21}/\Delta m^2_{32}|\ll 1$).  With NSI the problem is
intrinsically different, and requires a different approach, as
illustrated below. The analysis is carried out with $\theta_{13}=0$
for simplicity; corrections due to $\theta_{13}$
will be described  in sec. \ref{sub} and Appendix \ref{sect:app}. 

Let us start by considering the low energy part of the atmospheric
neutrino spectrum: $E \sim 0.1 - 1 $ GeV. Here we have $|\Delta | \gg
\sqrt{2}G_F n_e \gta \Delta_\odot $, and a two neutrino reduction is
rather simple: the observations can be described in terms of dominant
$\numu - \nutau$ (vacuum) oscillations driven by $\Delta$, with small
corrections due to $\nue \rightarrow \numu/\nutau$ oscillations driven
by the solar scale $\Delta_\odot$ and matter effects.  We will give
more details on these in Sect.~\ref{sub}.

At higher energy, $E \sim 1 - 5 $ GeV, we have $\sqrt{2}G_F n_e \sim
|\Delta| \gg \Delta_\odot$. We can neglect the smaller mass splitting,
and put $\Delta_\odot=0$.  In general, however, this approximations is
not enough to reduce to a two-neutrino problem, and so oscillations
cannot be studied analytically.
This follows from the fact that the mixing $\theta_{23}$ couples the
$\numu$ and $\nutau$ flavors, and the flavor-changing NSI term $\epet$
couples $\nutau$ with $\nue$.  This is an important difference with
respect to the case of SM interactions (or, more generally,
flavor-preserving interactions), where having $\epet=0$ allows to
decouple the $\nue$ state and reduce to a $\numu$-$\nutau$ system.

An important, nontrivial two-neutrino reduction is possible in the
\emph{highest energy limit}: $\sqrt{2}G_F n_e \gg |\Delta | \gg
\Delta_\odot $, which is well realized for $E \gta 10 $ GeV. As
mentioned in Sect.~\ref{minirev}, at these energies the observed signal
is just $\numu$ and $\barnumu$ disappearance, due to the absorption of
electrons.  Thus, the focus here is primarily on the conversion of
$\numu$ and $\barnumu$.

To see the two neutrino reduction, it convenient to introduce the
eigenvalues of $H_{mat}$:
\begin{widetext}
\beq
&&\lambda_{e '}=\frac{\sqrt{2}G_F n_e}{2}\left[
    1+\epee+\eptt+\sqrt{(1+\epee-\eptt)^2+4 |\epet|^2}
  \right]~,\nonumber \\ &&\lambda_{\mu '}=0~,\nonumber \\
  &&\lambda_{\tau'}=\frac{\sqrt{2}G_F n_e}{2}\left[
    1+\epee+\eptt-\sqrt{(1+\epee-\eptt)^2+4 |\epet|^2} \right]~.
\label{eig_he}
\eeq
\end{widetext}
and the matter angles $\beta$ and $\psi$:
\begin{eqnarray}
\tan 2\beta &=& 2 | \epet|/( 1+\epee -\eptt),\nonumber\\
2 \psi &=& Arg(\epet)~.
\label{beta_psi}
\end{eqnarray}

First, consider a situation in which both of the matter eigenvalues
dominate over the vacuum terms: $ |\lambda_{\tau'}|, |\lambda_{e '}|
\gg \Delta \gg \Delta_\odot $. In this case, the mixing $\numu$ in the
eigenstates of the Hamiltonian is suppressed by $\sim
\Delta/|\lambda_{e '}| \ll 1$. This means the muon neutrino will not
oscillate in the matter of the Earth, in conflict with the data. This
case then can be excluded with confidence.

Now, let us consider a very important case when the epsilons
compensate to a certain degree to give a \emph{hierarchical scheme} of
the type $ |\lambda_{e '}| \gg \Delta \sim |\lambda_{\tau'}| $ or $
|\lambda_{\tau'}| \gg \Delta \sim |\lambda_{e '}| $. This is the case
when the mention two-state reduction is realized. Indeed, the effects
of the matter interactions decouple one of the neutrino states, while
the vacuum term $\Delta$ drives the oscillations between the remaining
two.

Let us illustrate this for the situation $ |\lambda_{e '}| \gg \Delta
\sim |\lambda_{\tau'}| $, since it is smoothly connected to the
standard (no NSI) case and therefore appears more natural. Our results
can be easily generalized to the second scenario.

It is convenient to take the matter eigenstates:
\beq
&& \nue'=\cos\beta \nue + \sin \beta e^{-2 i \psi} \nutau~,\nonumber\\
&&\numu'=\numu~,\nonumber\\ && \nutau'=-\sin \beta e^{2 i \psi} \nue +
\cos \beta \nutau~.
\label{hameig}
\eeq
In the basis of the primed states, the Hamiltonian has the form:
\begin{widetext}
\be
H=\Delta \begin{pmatrix}
  -c^2_\beta + s^2_\beta c_{2\theta} + \lambda_{e '}/\Delta ~~~~ & s_\beta s_{2\theta} e^{-2 i \psi }  &  c_\beta s_\beta (1 + c_{2\theta}) e^{-2 i \psi } \\
  s_\beta s_{2\theta} e^{2 i \psi } &  - c_{2\theta}  &  s_{2\theta} c_\beta \\
  c_\beta s_\beta (1 + c_{2\theta}) e^{2 i \psi }   & s_{2\theta} c_\beta  & ~~~~~-s^2_\beta+ c^2_\beta c_{2\theta}+  \lambda_{\tau'}/\Delta \\
\end{pmatrix} ~.
\label{hamrot}
\ee
\end{widetext}
We see that the mixing of the state $\nue^\prime$ with the other
two is suppressed, being of the order $\sim \Delta/|\lambda_{e '}| \ll
1$.  Thus, this state decouples, and the problem reduces to $\numu
$-$\nutau^\prime$ oscillations described by the 2-3 block of the
Hamiltonian (\ref{hamrot}).  The unsuppressed oscillations of $\numu$
into a combination of $\nue$ and $\nutau$ can account for the observed
$\numu$ disappearance at high energy.

The two-state $\numu $-$\nutau^\prime$ system is quite simple to deal
with and indeed we may apply to it the known formalism for
two-neutrino oscillations in matter, (see e.g.
\cite{Mikheyev:1989dy,Kuo:1989qe}), with $\lambda_{\tau'}$ playing the
role of matter potential \footnote{The analogy with the standard
  description of neutrino propagation in matter is accurate if $\beta$
  is constant along the neutrino trajectory.  This is realized for NSI
  on electrons, or for NSI on quarks if the neutrons to protons ratio
  is constant. In other cases the dependence on time (distance) of the
  reduced 2$\times$2 Hamiltonian will be more complicated due to the
  t-dependence of $\beta$.}.
We find the effective mixing and mass splitting in matter:
\begin{eqnarray}
  &&\tan 2\theta_m = \frac{2 s_{2\theta} c_\beta}{c_{2\theta} (1+
    c^2_\beta) - s^2_\beta +\lambda_{\tau'}/\Delta}, \nonumber \\
  &&\Delta_{m} =
  \frac{\Delta}{2} \left[ \left(c_{2\theta} (1+ c^2_\beta)
      - s^2_\beta+\frac{\lambda_{\tau'}}{\Delta}\right)^2 +
    4s_{2\theta}^2 c_{\beta}^2\right]^{\frac{1}{2}},\;\;\;
  \label{eff_par}
\end{eqnarray}
and the equation for the oscillation probability in medium with
constant density:
\begin{equation}
  P(\numu\rightarrow \nutau^\prime)=\sin^2 2 \theta_m 
  \sin^2\left( \Delta_m L \right) ~.
  \label{p_hier}
\end{equation}
The expressions (\ref{eff_par}) give $\theta_m=\theta$ and
$\Delta_{m}=\Delta$ if $\epet=\eptt=0$.  Notice that along the
direction $\epet=0$ the matter eigenstates (\ref{hameig}) coincide with the flavor
ones, and we recover the 2$\times$2 problem of Refs.
\cite{Fornengo:2001pm,Gonzalez-Garcia:2004wg}, in which $\nue$ is
decoupled, resulting in no sensitivity to $\epee$.

The properties of the probability $P(\numu\rightarrow \nutau^\prime)$
in Eq.  (\ref{p_hier}) follow those of the MSW effect, with the
possibility of resonant amplification of the oscillation amplitude
($\sin 2\theta_m=1$) when the condition $c_{2\theta} (1+ c^2_\beta) -
s^2_\beta +\lambda_{\tau'}/\Delta=0$ is realized.  It is worth
noticing that $\theta_m$ and $\Delta_m$, and thus the probability
$P(\numu\rightarrow \nutau^\prime)$, do not depend on the phase of the
$e-\tau$ NSI term, $\psi$.
\\

Given its particular relevance for the analysis of atmospheric
neutrinos, we now comment on the limit of small $\lambda_{\tau'}$,
i.e.  $|\lambda_{e '}| \gg \Delta \gg |\lambda_{\tau'}| $. This
corresponds to a parabola in the space of the NSI couplings for fixed
$\epee$: 
\begin{equation} 
  \eptt \sim |\epet|^2/(1+\epee)~.
  \label{parab}
\end{equation} 
If we take $\lambda_{\tau'}=0$ in Eqs. (\ref{eff_par}) and
(\ref{p_hier}), we see that here the phase of the $\numu
$-$\nutau^\prime$ oscillations has same dependence on the product
$\Delta \cdot L$ of vacuum oscillations, while at the same time the
interaction with matter is far from negligible, as it enters the
probability through the angle $ \beta$. We also notice that the conversion does not depend on
the sign of $\Delta$ (mass hierarchy). It is also independent of the
overall sign of the matter Hamiltonian, therefore neutrinos and
antineutrinos have the same oscillation probability.

\subsection{Expected sensitivity} \label{dom}
What values of $\epee,\epet,\eptt $ are compatible with the
atmospheric and K2K neutrino data? And, how does the presence of NSI
change the allowed region in the space of $\theta_{23}$ and $\Delta
m^2_{32}$?  While to find the allowed region in the five-dimensional
parameter space requires a numerical scan, a good part of the relevant
features of this region can be understood from analytical arguments.
Here we illustrate those. For simplicity, we first consider how the
atmospheric neutrino data put constraints on NSI, and successively
analyze how the K2K signal contributes to tighten those limits.

Let us consider several different regimes.

\underline{FP couplings only: $\epet=0$}.  Here we expect that
practically any value of $\epee$ will be compatible with the data, due
to the decoupling of $\nue$, as explained in Sect.~\ref{2nured}.  In
contrast, the data strongly constrain $\eptt$.  Indeed, $\eptt$
influences the oscillations in the $\numu-\nutau$ sector, with
possible conflict with observations. More specifically, an upper bound
on $\eptt$ comes from requiring that the $\numu \leftrightarrow
\nutau$ oscillation amplitude remains maximal over a large interval of
energy, $E \sim 10^{-1} - 10^2$ GeV, as indicated by the data.  This
can only be realized if $\theta_{23}$ is maximal and if the matter
term remains subdominant to the vacuum ones even at the highest
energies\footnote{ One could devise a resonant MSW-like solution,
  where $\theta_{23}$ is away from maximal mixing but the mixing in
  matter is maximal at $E\sim E_0$ in one channel (neutrinos or
  antineutrinos, but not both due to the different sign of the matter
  potential) due to the cancellation of vacuum and matter terms:
  $\sqrt{2}\eptt G_F n_e - \Delta m^2/(2 E_0)\sim 0$. This scenario is
  not viable due to the suppression of mixing in the other channel and
  also because it would poorly fit the sub-GeV data, which require
  maximal mixing.}.  For neutrinos going through the center of the
Earth, the highest energy at which an oscillation minimum occurs in
the standard case is around
\begin{equation}
  \label{eq:E0}
  E_0 \sim 20-30 \mbox{ GeV}.
\end{equation}
If we require $\sqrt{2}\eptt G_F n_e\lta \Delta m^2_{32}/(2 E_0)$, and
use $\Delta m^2_{32} \simeq 2.5 \cdot 10^{-3}~{\rm eV^2}$, we find the
bound $\eptt\lesssim 0.2$. This agrees well with numerical results
\cite{Friedland:2004ah} (see sec. \ref{sect:num}). We also expect no
modifications of the allowed region of $\theta_{23}$ and $\Delta
m^2_{32}$ with respect to the standard case.

\underline{Both FP and FC couplings: no cancellations}.  In the
presence of flavor changing interactions, $\epet\neq 0$, a more
general bound on the NSI follows from the analysis of
Sect.~\ref{2nured}.  Let us begin with the ``generic'' scenario in
which $|\lambda_{\tau'}|, |\lambda_{e '}| \gg \Delta $ for $E \sim
E_0$. This case is clearly excluded. Indeed, in this case the $\numu$
mixing is suppressed (see Sect.~\ref{2nured}), and thus the muon
neutrino flux remains unoscillated, in clear conflict with the data.
This scenario yields what could be called a ``generic'' bound on the
epsilons: for example, we get $|\epet|\lta 0.5$ when $\epee \sim 0$.
The bound depends on $\epee$ in a way that will be described later.

\underline{Both FP and FC couplings: hierarchical scenario}.  We now
analyze the ``hierarchical'' scenario, where $|\lambda_{e '}| \gg
\Delta \sim |\lambda_{\tau'}| $ for $E\sim E_0$.  This configuration
turns out to fit the data even for rather large values of $\epet$ and
$\eptt$.  To understand the reason, we can first focus on the limiting
case $\lambda_{\tau '}= 0$, corresponding to the parabolic direction
in Eq. (\ref{parab}).  As explained in Sect.~\ref{2nured}, in this
circumstance the muon neutrino evolution allows a two-state
$\numu-\nutau^\prime$ reduction and the resulting disappearance of
$\numu$ has the all features of vacuum oscillations that are known to
fit the data, namely the same $L/E$ dependence and equal survival
probabilities for neutrinos and antineutrinos.  The fact that here
$\numu$ oscillates into a combination of $\nutau$ and $\nue$, and not
into pure $\nutau$ as in the standard scenario, is inconsequential,
since there are no e-like data available at $E\gta 10$ GeV, as
mentioned in sec.  \ref{minirev}. Moreover, at $E \ll E_0$ vacuum
terms are dominant and so vacuum oscillation features are recovered in
the lower energy data samples.  The result is an allowed region in the
space of the NSI couplings centered along the parabola (\ref{parab}).

Let us study the allowed region of the hierarchical case, and in
particular its width and extent along the parabolic direction.
The width of the region is given by the condition 
$|\lambda_{\tau'}|<\Delta m^2_{32}/(2 E_0)$, or, numerically:
\begin{equation}
  \label{eq:width}
  |1+\epee+\eptt-\sqrt{(1+\epee-\eptt)^2+4 |\epet|^2}|\lesssim 0.4~.
\end{equation}
To determine the extent of the region, more complicated considerations
are necessary.  As a first step, let us address the question of how
the NSI change the vacuum oscillations parameters reconstructed from
the data.  We expect that, if the multi-GeV and through-going muon
events have a significant weight in the global fit to the data, for a
given set of NSI in the region (\ref{eq:width}) the allowed region in
the space of $\theta_{23}$ and $\Delta m^2_{32}$ changes with respect
to the standard case.  Indeed, if NSI are present, but not included in
the data analysis, a fit of the highest energy atmospheric data, i.e.
the through-going muon ones, would give $\Delta m^2_{m} $ and
$\theta_m $ instead of the corresponding vacuum quantities.  If we
require that the measured $\theta_m$ is maximal, as favored by the
data, Eq. (\ref{eff_par}) gives: 
\begin{eqnarray} 
  &&\cos 2\theta_{23} \simeq s^2_\beta /(1+ c^2_\beta)~, \label{vac_mix} \\ 
  &&\Delta m^2_{32} \simeq \Delta m^2_m(1+\cos^{-2}\beta)/2~,
  \label{vac_dm2}
\end{eqnarray}
where $\beta$ is the rotation angle between the NSI eigenbasis
and the flavor basis, Eq. (\ref{beta_psi}).  Interestingly, Eqs.
(\ref{vac_mix}) and (\ref{vac_dm2}) tell us that the vacuum mixing
$\theta_{23}$ would {\it not} be maximal and that $\Delta m^2_{32}$
would be larger than the measured value. The fit to the high-energy
dataset can be achieved, but only \emph{at the expense of modifying
  the vacuum oscillation parameters}.
At the same time, the low-energy (sub-GeV) dataset still has
negligible matter effect and hence is best fit by maximal
$\theta_{23}$.  Thus, for sufficiently large NSI, there will be
\emph{a tension} between the values of the oscillation parameters
preferred by the low- and high-energy datasets. Avoiding this
tension leads to a constraint on $\beta$, which in turn translates
into a constraint on the epsilons.

More specifically, one should impose that the angle given by Eq.
(\ref{vac_mix}) is larger than the minimum value of $\theta_{23}$ --
let us call it $\theta_{min}$ -- allowed by the low energy sample.
This gives:
\begin{equation}
\cos^2 \beta \gta \tan^2 \theta_{min}~,
\label{alexcond}
\end{equation}
which will cut off the potentially infinite parabola (\ref{eq:width})
down to the shape of a smile (see sec.  \ref{sect:num}).
Fits to the sub-GeV data sample give $\theta_{min} \simeq 0.52$
\cite{Gonzalez-Garcia:2004cu}. Using this value we get $\cos^2 \beta
\gta 0.3$; this result is relaxed if we allow $\theta_m$ to be
slightly away from $\pi/4$ in the derivation of Eq. (\ref{vac_mix}).

Similarly, we have to require that the right hand side of Eq.
(\ref{vac_dm2}) does not exceed the maximal mass splitting $\Delta
m^2_{max}$ allowed by the sub-GeV data, and obtain: 
\be 
\cos^2 \beta \geq 
\left[\frac{2 \Delta m^2_{max}}{\Delta m^2_m}-1 \right]^{-1}~.
\label{cecicond}
\ee 
For $\Delta m^2_{max}= 5.0 \cdot 10^{-3}~{\rm eV^2} $
\cite{Gonzalez-Garcia:2004cu} and $\Delta m^2_m\simeq 2.5 \cdot
10^{-3}~{\rm eV^2} $ Eq.~(\ref{cecicond}) gives $\cos^2 \beta \gta
0.3$, comparable to what given by Eq.~(\ref{alexcond}).

While Eq.~(\ref{eq:width}) was given in our earlier work
\cite{Friedland:2004ah}, the results in Eqs.~(\ref{alexcond}) and
(\ref{cecicond}) are presented here for the first time.

Let us check if the constraints in Eqs.~(\ref{alexcond}) and
(\ref{cecicond}) become stronger if we combine the atmospheric
neutrino data with those from K2K.  For the neutrino energies used at
K2K and $\epsilon_{\alpha \beta}\lta 1$, matter effects are
negligible, and therefore K2K measures the vacuum parameters
$\theta_{23}$ and $|\Delta m^2_{32}|$.  From the K2K limit on
$\theta_{23}$ we get the condition to avoid the tension between K2K
and atmospheric data. The analogue of Eq.~(\ref{alexcond}) for K2K
turns out to be looser than that from sub-GeV atmospheric events,
while the condition on the mass splitting, Eq.~(\ref{cecicond}) is
important: using the K2K limit, $\Delta m^2 \lesssim 4.0 \cdot
10^{-3}~{\rm eV^2}$ \cite{Aliu:2004sq}, we find $\cos^2 \beta \gta
0.45$. Thus, the present K2K data constrains NSI by limiting the
allowed range of $\Delta m^2$.

If the opposite hierarchy of eigenvalues is realized, $
|\lambda_{\tau'}| \gg \Delta \sim |\lambda_{e'}| $, similar
considerations to those above apply.  The allowed region in the space
of the NSI couplings is now given by the requirement $|\lambda_{e'}|
\lta \Delta m^2_{32}/(2 E_0)$, or, numerically:
\begin{equation}
  \label{eq:width_neg}
  |1+\epee+\eptt+\sqrt{(1+\epee-\eptt)^2+4 |\epet|^2}|\lesssim 0.4~.
\end{equation}
This condition can be satisfied for $\epee\lesssim -1$. It describes a
region centered around the parabola (\ref{parab}) oriented in
the negative $\eptt$ direction.  Physically, this means that the
contribution of $\epee$ overcompensates the standard matter term and
changes the sign of the $e$-$e$ entry of the matter Hamiltonian.  This
is somewhat extreme but still compatible with accelerator limits (Eq.
(\ref{epbounds})) and with the combination of solar and KamLAND data,
provided that a certain combination of the other parameters (NSI and
vacuum ones) is realized \cite{Guzzo:2004ue,Miranda:2004nb}.

As we decrease $\epee$ from $\epee \sim 0$ down to $\epee \sim -2$ a
transition from one limiting case ($ |\lambda_{e'}| \gg \Delta \sim
|\lambda_{\tau'}| $) to the other ($ |\lambda_{\tau '}| \gg \Delta
\sim |\lambda_{e'}| $) occurs.  We expect this transition to be
continuous, meaning that for each fixed value of $\epee$ in this
interval a region in the $\epet$-$\eptt$ plane that is compatible with
the data exists and varies smoothly with $\epee$.  This conclusion is
justified by our earlier comment that atmospheric neutrino
oscillations are insensitive to NSI in the direction $\epet=\eptt=0$.
Numerical results indeed confirm the existence of such a continuous
transition (see Sect.~\ref{sect:num}).

%
\section{Atmospheric neutrinos with NSI: numerical analysis}
\label{sect:num}

\subsection{Combined analysis of atmospheric and K2K data}
\label{atmcode}

\begin{figure*}[htbp]
\centering
\includegraphics[height=0.9\textheight]{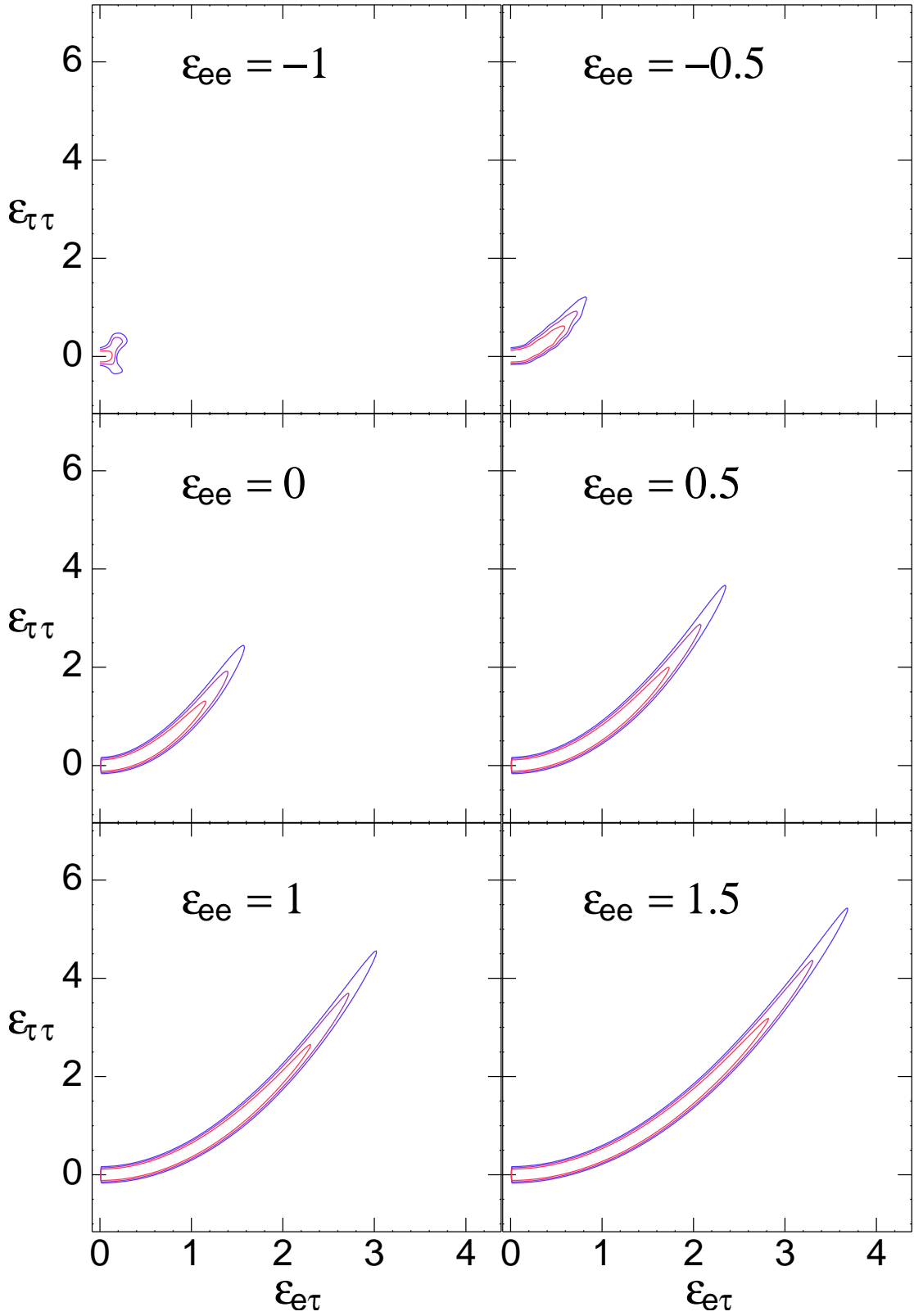}
\caption{Sections of the function $\chi^2(\epee,\eptt,\epet)$ along
  the planes of fixed $\epee$ (numbers in the Figure), from a combined
  fit of the atmospheric and K2K data. The epsilons shown are per
  electron. The vacuum parameters $\theta_{23}$ and $\Delta m^2_{32}$
  are marginalized over. The latter is taken to be negative (inverted
  mass hierarchy). We have also used $\theta_{13}=0$ and neglected
  effects of the smaller mass splitting, $\Delta m^2_{21}=0$. The
  contours, from the inner to the outer, correspond to $95 \%, 99\%$
  and $3\sigma$ C.L. Only the positive $\epet$ semi-plane is shown in
  each case for visual convenience; the negative $\epet$ semi-plane is
  symmetric. More generally, the results are independent of the
  complex phase of $\epet$.}
\label{atmk2k6panels}
\end{figure*}

We performed a quantitative analysis of the atmospheric neutrino
data with five parameters: two ``vacuum" ones, $(\Delta
m^2_{32},\theta_{23})$, and three NSI quantities
$(\epee,\epet,\eptt)$, where $\epet$ has been treated as real for
simplicity. The parameter space was scanned and a goodness-of-fit
analysis was performed for each grid point.

In the analysis, we have used two types of codes. For many of the
preliminary investigations we used our own code, which made several
simplifying assumptions, but was designed to capture the relevant
physical features of the atmospheric neutrinos in different energy
ranges. For the final fits, we used a binary of the atmospheric
neutrino program kindly provided to us by Michele Maltoni (SUNY, Stony
Brook). This binary is essentially the same program used in our
earlier paper \cite{Friedland:2004ah}. It uses the complete 1489-day
charged current Super-Kamiokande phase I data set~\cite{Hayato:2003},
including the $e$-like and $\mu$-like data samples of sub- and
multi-GeV contained events (each grouped into 10 bins in zenith angle)
as well as the stopping (5 angular bins) and through-going (10 angular
bins) upgoing muon data events. This amounts to a total of 55 data
points. For the calculation of the expected rates the code adopts the
three-dimensional atmospheric neutrino fluxes given in
Ref.~\cite{Honda:2004yz}. The statistical analysis of the data follows
the appendix of Ref.~\cite{Gonzalez-Garcia:2004wg}. The binary
underwent extensive testing in the course of this project and the
feedback was reported back to the author.

We have included the results of the K2K data analysis in our study, by
adding our atmospheric $\chi^2$ to the K2K $\chi^2$ in the space of
$\theta_{23}$ and $\Delta m^2_{32}$. The latter was provided by the
K2K collaboration in tabular form, and refers to the published
analysis of Ref. \cite{Aliu:2004sq} (Fig. 4 there).

It is important to point out that in our analysis we use only NSI
\emph{propagation} effects. The \emph{detection} effects are
purposefully left out. The reason for this is that the changes in the
detection cross sections due to NSI do not uniquely follow from the
propagation effects, but depend on additional parameters and
assumptions. As a result, the detection effects can vary
significantly, from large to unobservable, depending on the underlying
model of the NSI.

\subsubsection{NSI and detection effects}

Let us give a detailed argument for why possible NSI effects on the
detection cross sections are not directly related to the propagation
effects. A reader primarily interested in the results of our numerical
analysis may wish to skip to Sect~\ref{mainres}.

Consider how the Super-Kamiokande collaboration extracts NC
information from their data. One method is to analyze a multi-ring
dataset specifically enriched with NC events through a careful
sequence of cuts \cite{Fukuda:2000np,phd1}. Another method is to
compare the rate of single $\pi^0$ events, most of which are produced
in NC interactions, to the muon rate \cite{phd2}. In principle, only
the collaboration can reliably model these data. Nevertheless, in what
follows we set this consideration aside and ask: Can one in principle
use these data to improve the bounds on NSI derived from propagation
effects?

\underline{The multi-ring dataset}. The main contribution to this
dataset is from multi-pion production in the energy range $0.5 - 10$
GeV \cite{phd1}. While the exact expression for the cross-section is
quite complicated (SK uses a semi-empirical formula), a qualitative
estimate may be obtained by assuming deep inelastic neutrino-parton
scattering. In this case, the cross-section depends on the squares of
the left- and right-handed couplings, $g_L^2$ and $g_R^2$
\cite{Fukugita:2003en}. It is immediately obvious that the result
depends on how the NSI effects are distributed between the $u$ and $d$
quarks. The value of the axial coupling, $g_A=g_L-g_R$, is also
crucial (the refraction effects fix only the vector part,
$g_V=g_L+g_R$). For example, for the point used in
\cite{Friedland:2004pp}, splitting NSI evenly between $g_L$ and $g_R$
and assuming the standard value for $g_A$, we find an increase in the
cross-section of only 20\%. The effect becomes even smaller once the
axial coupling is adjusted to compensate the increase given by the
vector part. Hence, no rigorous constraint can be obtained from this
sample.

\underline{The single $\pi^0$ sample}. The main contribution to the
single $\pi^0$ sample is from incoherent single pion production,
followed by coherent single pion production \cite{phd2}. The first
process is dominated by the $\Delta$ resonance and is largely
controlled by the size of the axial coupling \cite{Fukugita:2003en}.
The second one is also axial: the neutral current creates a pion that
scatters on the nucleus \cite{{rein}}. Hence, this sample would not
constrain the vector interaction that is responsible for the modified
matter effect.  Also, it should be mentioned that the statistics in
this sample is not sufficient to even separate the active-active from
active-sterile scenarios. (The latter is disfavored by only $1.3
\sigma$, see \cite{phd2}.)

Lastly, a simple observation is that if the NSI are assigned to
electrons, they have no effect on the NC event rate in the detector,
since the latter is dominated by scattering on nuclei.  This makes
especially clear that the propagation and detection effects need not
be correlated. Only if strong model-dependent assumptions are made,
like in the case of a sterile neutrino, can the two be used together
to exclude an oscillation scenario.

We emphasize that it is important for Super-Kamiokande and other
neutrino oscillation experiments to be looking for any anomalous NC
signal: an observation of such signal would imply the discovery of
NSI. At the same time, as the preceding examples show, a lack of such
anomalous signal would not guarantee there are no NSI propagation
effects.

\subsection{Main results: NSI, mixing and mass splitting}
\label{mainres}

Our main result involves a scan of the
$(\epsilon_{ee},\epsilon_{e\tau},\epsilon_{\tau\tau},\Delta
m^2_{32},\theta_{23})$ space for real $\epet$, inverted mass hierarchy
($ \Delta m^2_{32}<0$), $\theta_{13}=0$ and $\Delta m^2_{21}=0$.
Variations of the three latter parameters represent subdominant
effects; the cases of different mass hierarchy and nonzero
$\theta_{13}$ are treated later, in Sect.~\ref{sub}.

\begin{figure}[htbp]
\centering
\includegraphics[width=0.45\textwidth]{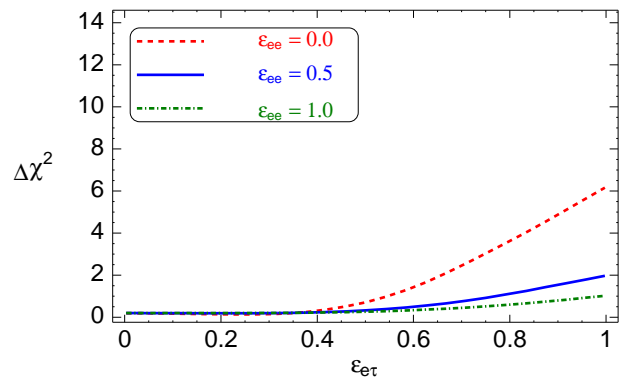}
\caption{ The difference $\Delta \chi^2$ (where $\chi^2$ is the same
function of Fig. \protect\ref{atmk2k6panels}) plotted along the
parabolic direction of Eq. (\protect\ref{parab}), for three fixed
values of $\epee$
(see legend). }

\label{chi2profilesinv}
\end{figure}

\begin{figure}[htbp]
\centering
\includegraphics[width=0.45\textwidth]{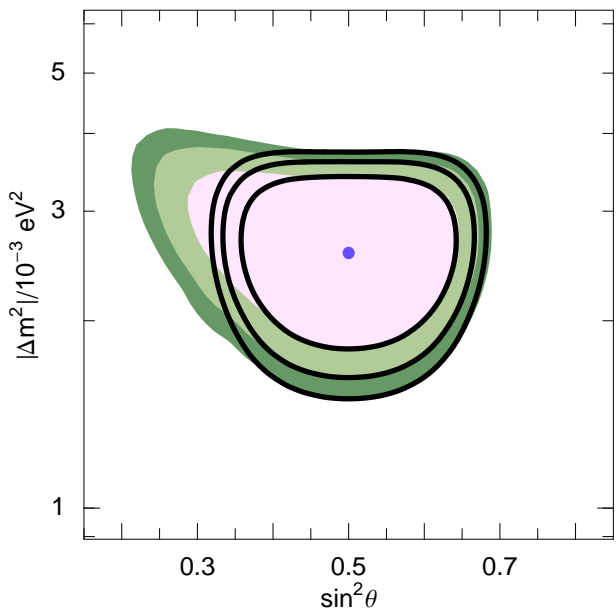}
\caption{The shaded areas represent the regions in the space of
  $|\Delta m^2_{31}|$ and $\sin^2 \theta_{23}$ allowed by the
  atmospheric neutrino and K2K data, obtained upon marginalizing over
  the NSI parameters. The regions correspond to the 95\%, 99\% and
  $3\sigma$ confidence levels for 2 d.o.f. (innermost to outermost).
  We have marginalized also over the sign of $\Delta m^2_{31}$ and
  took $-1.6\leq \epee \leq 1.6$, motivated by one of the accelerator
  bounds (see text). The corresponding regions obtained with purely
  standard interactions are shown for comparison.}
\label{finalanswer}
\end{figure}

\begin{figure}[htbp]
\centering
\includegraphics[width=0.45\textwidth]{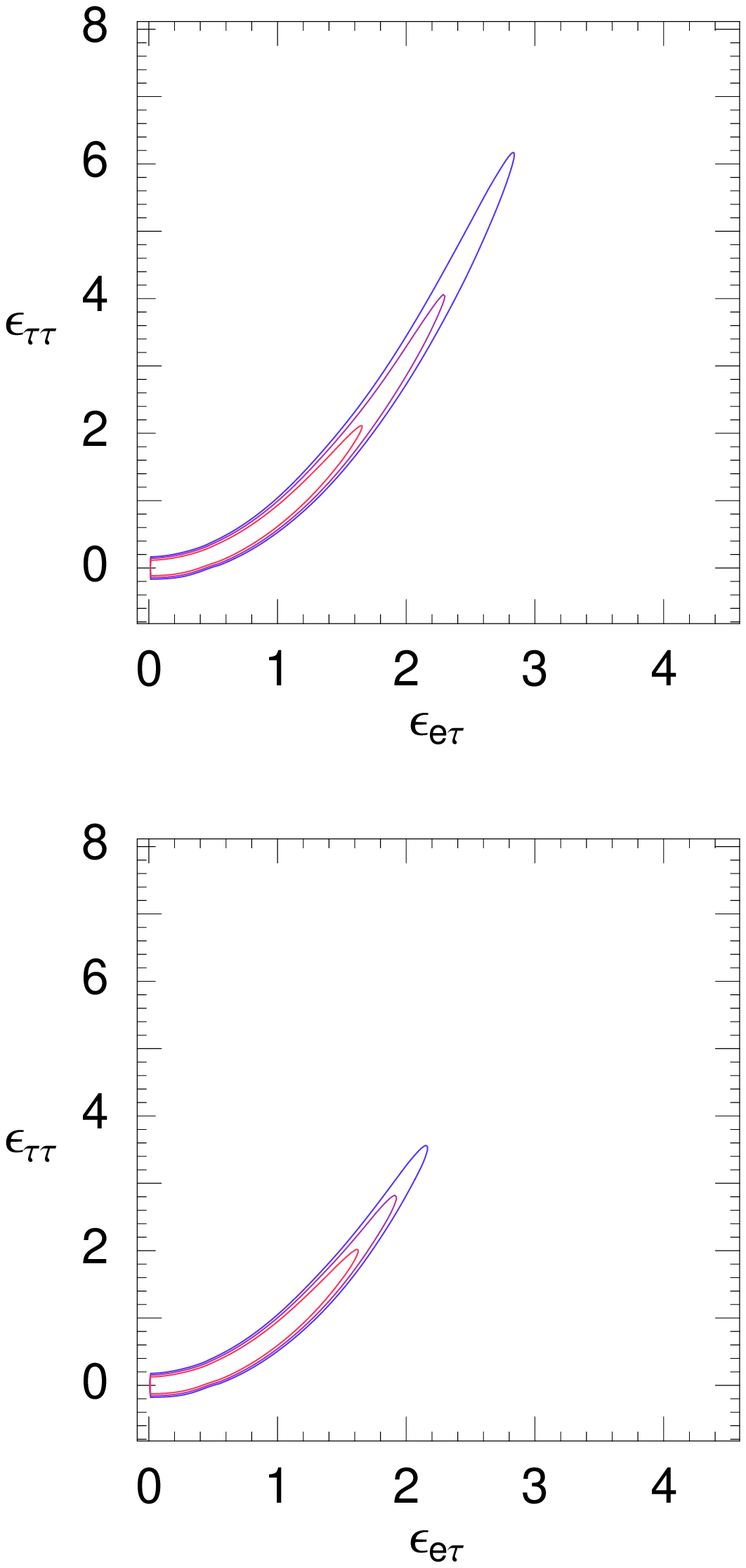}
\caption{
A section of the function $\chi^2(\epee,\eptt,\epet)$ along
the plane $\epee=0.3$, from a fit to atmospheric data only (upper) and
the combined fit of atmospheric and K2K data (lower). All the
considerations made for Fig. \protect\ref{atmk2k6panels} apply here.}
\label{k2keffectee03}
\end{figure}

\begin{figure*}[htbp]
\centering
\includegraphics[width=0.9\textwidth]{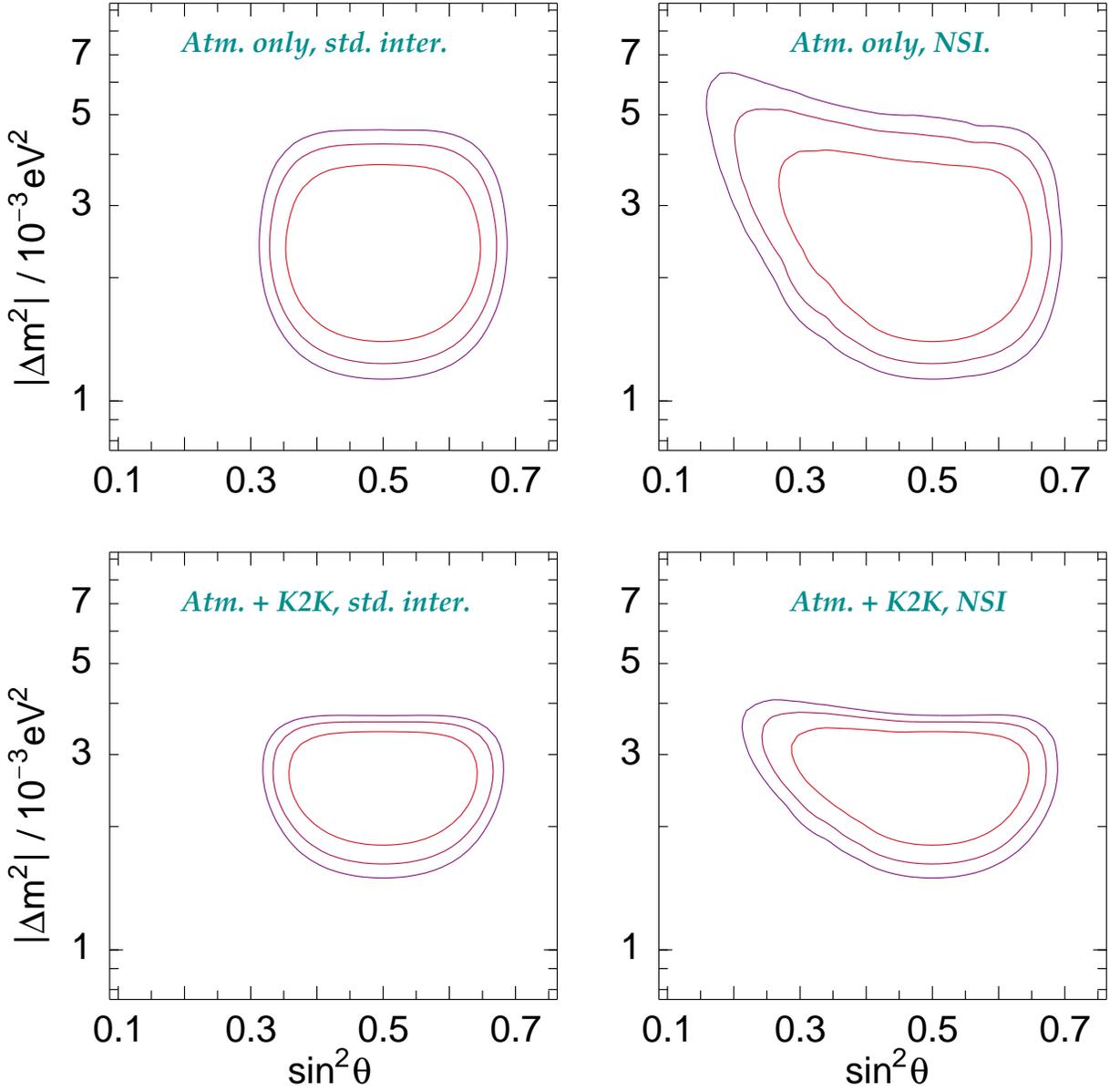}
\caption{
A more detailed version of Fig.~\protect\ref{finalanswer}, where
the results of fitting the atmospheric data only and atmospheric+K2K
are given for both the standard case and the case with NSI.}
\label{margin_ee16_k2k}
\end{figure*}

The scan yields a five-dimensional allowed region. Various projections
of this region are described next, in Figs.
\ref{atmk2k6panels}-\ref{margin_ee16_k2k}.

The first result is the region allowed by the data in the space of the
NSI. It was found by marginalizing the $\chi^2$ function over the
vacuum oscillation parameters.  The resulting function,
$\chi^2(\epee,\epet,\eptt)$, gives a 3-dimensional allowed region,
two-dimensional $(\epet,\eptt)$ sections of which are shown in Fig.
\ref{atmk2k6panels} (the values of $\epee$ for each section are shown
in the Figure). As the region is invariant under $\epet
\leftrightarrow -\epet$, only the positive $\epet$ halves are shown for
each section.  The symmetry in the sign (phase) of $\epet$ is
understood in terms of the oscillations probabilities, that were shown
to depend on the absolute value of $\epet$ only (see sec.  \ref{dom}).

The curves in the Figure are isocontours of $\Delta
\chi^2\equiv\chi^2-\chi^2_{min}= 7.81,11.35,14.16$ (from the inner to
the outer), corresponding to $95\%,99\%$ and $3\sigma$ C.L. The
parabolic flat direction in the function $\chi^2(\epee,\epet,\eptt)$
predicted by Eq.~(\ref{parab}) is clearly seen.  This direction is
essentially determined by the atmospheric data, and is not altered
significantly by the contribution of  K2K (which, however, changes
the extent of the region, as will be explained). The width of the
parabolic region matches well the condition on the matter eigenvalues,
Eq. (\ref{eq:width}), as was shown explicitly in Fig.~1 of ref.
\cite{Friedland:2004ah}.  The points where the region (at a given
C.L.) ends along the parabola well follow isocontours of the angle
$\beta$ (specifically, $\cos^2 \beta\simeq 0.30,0.35,0.42$ for the
three confidence level contours in the figure), as expected from the
``cutoff" conditions given by the low energy atmospheric sample and by
the K2K results, Eqs.~(\ref{alexcond}) and (\ref{cecicond}).

Along the parabola, the function $\chi^2(\epee,\epet,\eptt)$ varies
very slowly near the origin, and starts to increase appreciably only
at $\epsilon_{e\tau}$ of about 0.5 or so.  For example, we have:
$\chi^2(0,0,0)=148.11$ (no NSI) and $\chi^2(0.73,0.35, 0.07)=148.07$.
The latter happens to be the absolute minimum, $\chi^2_{min}$, but
clearly has practically the same goodness of fit as the origin.
The curves Fig.~\ref{chi2profilesinv} show how  $\Delta \chi^2$ varies with
$\epet$ along the parabola (\ref{parab}) for three fixed values of $\epee$.

As a side comment, we notice that the agreement with Eqs.
(\ref{parab}) and (\ref{eq:width}) becomes worse with the decrease of
$\epee$. This makes sense because as $\epee$ approaches $-1$, the two
matter eigenvalues $\lambda_{e^\prime}$ and $\lambda_{\tau^\prime}$
have comparable size, thus breaking the approximation of hierarchy of
the eigenvalues used in our derivations (sec. \ref{2nured}).  Notice
that the panel with $\epee=-1$ shows a hint of transition from an
upward to a downward parabolic region. This transition is expected to
happen with the change of sign of the term $1+\epee$ in the matter
Hamiltonian.

The second result is the allowed region in the space of $\theta_{23}$
and $|\Delta m^2_{31}|$, obtained by marginalizing over the NSI
parameters.  This region is shown in Fig. \ref{finalanswer}.  In the
marginalization procedure we have restricted $\epee$ in the interval
$-1.6 \leq \epee \leq 1.6$. This serves as an example, and corresponds
to the CHARM bound if the NSI are present exclusively as
flavor-preserving interactions of $\nue$ on the right-handed down
quark \cite{Davidson:2003ha} \footnote{A completely rigorous way to
  incorporate the CHARM bound would be to reanalyze the CHARM results
  with all the relevant NSI couplings in the $e-\tau$ sector
  simultaneously, and make a global fit of atmospheric, K2K and
  accelerator data. This is beyond the scope of this work. }.  We have
marginalized also over the sign of $\Delta m^2_{31}$, thus including
both mass hierarchies.  Expectedly, with respect to the standard case,
the allowed region is bigger, and extends to smaller mixing ($\sin^2
\theta_{23} \simeq 0.2$ instead then $\sin^2 \theta_{23} \simeq 0.31$)
and slightly larger $|\Delta m^2_{31}|$, in agreement with the
analytic considerations (sec. \ref{sect:analyt}).  The absolute
minimum lies at $|\Delta m^2_{31}|=2.6\times 10^{-3}$ eV$^2$, $\sin^2 \theta=0.5$.

We find that the contours in the figure practically coincide with
those obtained for normal hierarchy with $\epee$ fixed at the upper
limit, $\epee=1.6$.  Since this upper limit is determined by
accelerator experiments, improvements of the accelerator NSI bounds would
lead to a better knowledge of the oscillation parameters.

In the Table \ref{tabmassmix} we list the intervals allowed, at
different confidence levels, for each vacuum parameter after
marginalizing over all the other quantities.  They exhibit features
analogous to those of fig. \ref{finalanswer}.  The results for the
standard case are given, too, for comparison.

The K2K results play an important role in restricting both the region
of the oscillations parameters and that of the NSI couplings. This is
shown in Figs. \ref{k2keffectee03} and \ref{margin_ee16_k2k}.  Figure
\ref{k2keffectee03} shows a representative section of
$\chi^2(\epee,\eptt,\epet)$ (the same function as in Fig.
\ref{atmk2k6panels}) along the plane $\epee=0.3$. The lower panel
refers to the full atmospheric+K2K fit (like all the results in Figs.
\ref{atmk2k6panels}-\ref{finalanswer}), while the upper one is
obtained with atmospheric data only.  The reduction due to K2K is
evident: for example, the edge of the $3\sigma$ contour changes from
$(\epet,\eptt)\simeq (2.8,6.0)$ to $(\epet,\eptt)\simeq (2.2,3.8)$.
The effect is smaller for smaller NSI, as one can see that the $95\%$
C.L. contour is practically unchanged between the two panels.

Fig. \ref{margin_ee16_k2k} is a series of variations of
Fig. \ref{finalanswer}: the results of the fit in the plane $|\Delta
m^2_{31}|$-$\sin^2\theta_{23}$ are shown with and without NSI, each
with and without the inclusion of the K2K results. We see that with
NSI the K2K data contribute to restrict the parameters, especially in
the region of large $|\Delta m^2_{31}|$ and small mixing.  The
$3\sigma$  contour of this region is moved from
$(\sin^2\theta_{23},|\Delta m^2_{31}|)\simeq (0.15,6 \cdot
10^{-3}~{\rm eV^2})$ to $(\sin^2\theta_{23},|\Delta m^2_{31}|)\simeq
(0.21,4 \cdot 10^{-3}~{\rm eV^2})$. We also observe a restriction of
$|\Delta m^2_{31}|$ from below, which is essentially the same in the
cases with and without NSI.


\subsection{Subdominant effects: mass hierarchy, $\theta_{13}$,
  three neutrino corrections} \label{sub}

\begin{figure*}[htbp]
\centering
\includegraphics[height=0.9\textheight]{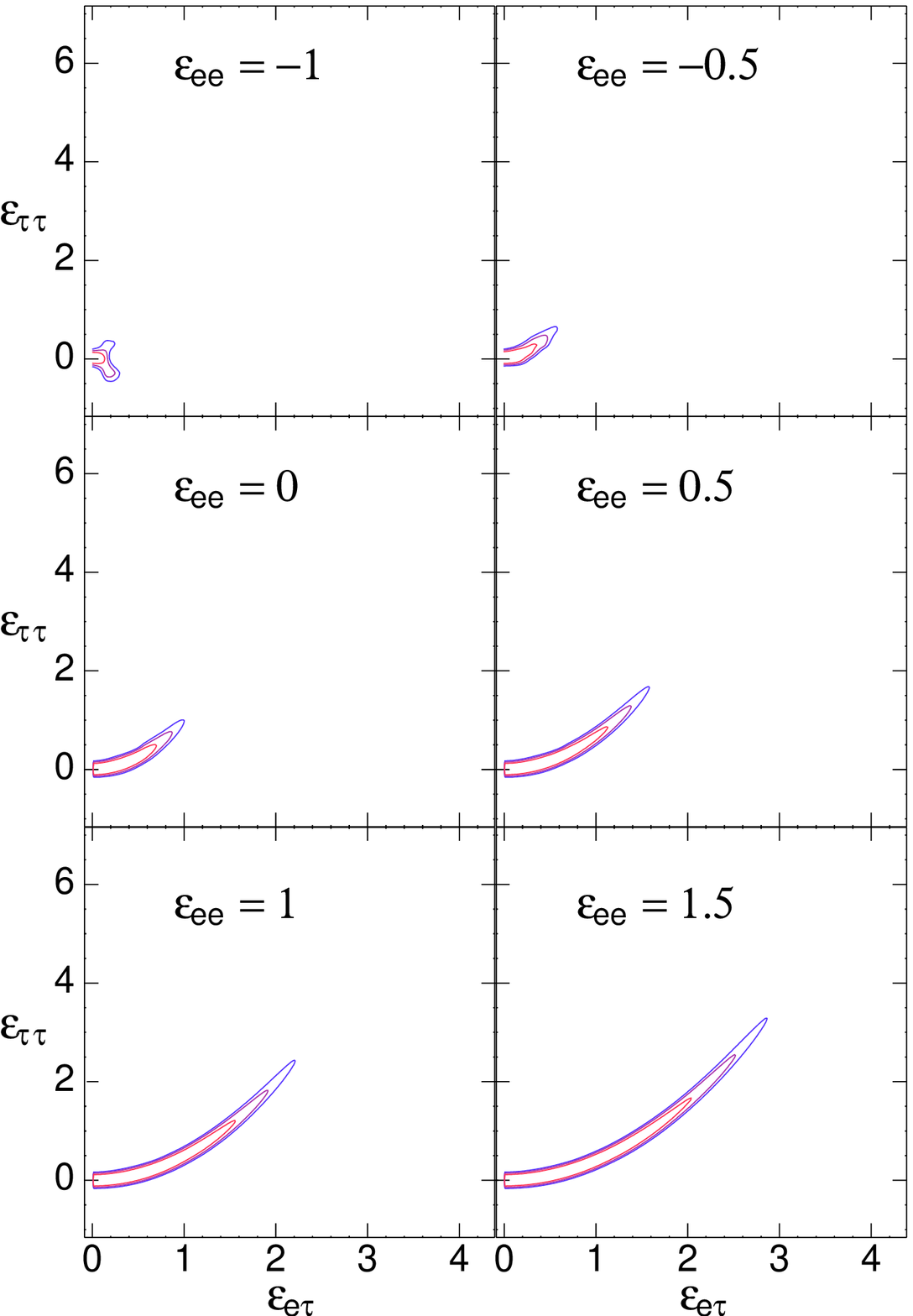}
\caption{The same as Fig. \protect\ref{atmk2k6panels} for normal mass
  hierarchy.}
\label{atmk2k6panelsnormal}
\end{figure*}

\begin{figure}[htbp]
  \centering
  \includegraphics[width=0.47\textwidth]{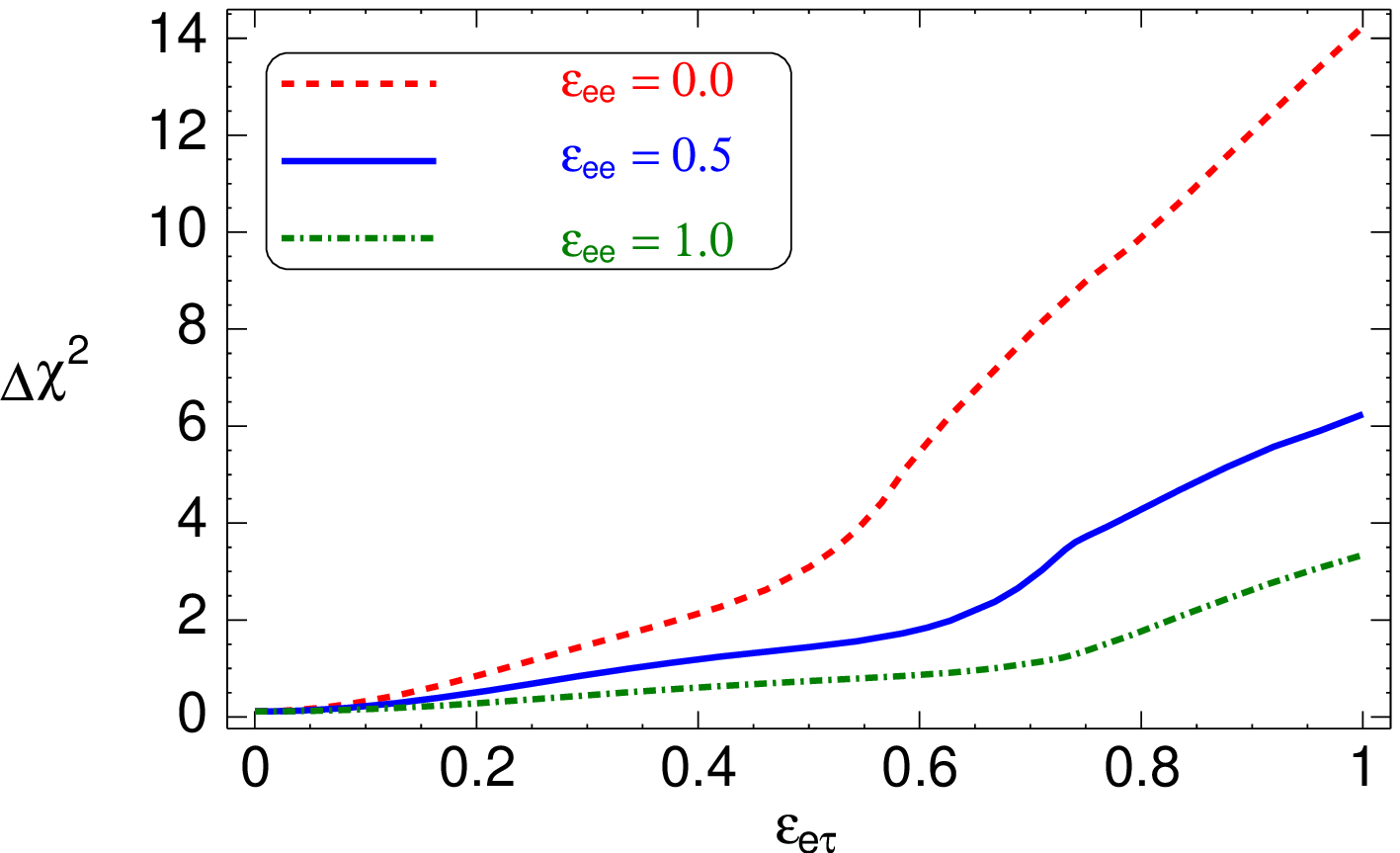}
\caption{ The same as Fig. \ref{chi2profilesinv} for normal mass hierarchy.}
\label{chi2profilesnorm}
\end{figure}

Here we generalize our results to include subdominant effects, namely
the effect of the sign of the mass hierarchy, of a non-zero
$\theta_{13}$ and of the 1-2, ``solar'', oscillation parameters.  

Fig.~\ref{atmk2k6panelsnormal} shows the allowed region in the space
of the NSI parameters for normal mass hierarchy.  Analogously to the
case of inverted hierarchy shown in Fig.~\ref{atmk2k6panels}, the
region follows the parabola (\ref{parab}) and its endpoints follow
isocontours of $\beta$.  The main difference is that for normal
hierarchy these isocontours are more restricted, {\it i.e.}, the
allowed range of NSI is smaller. Along the parabola, the $\chi^2$
grows faster with the epsilons for normal hierarchy, as shown in Fig.
\ref{chi2profilesnorm}.  For example, if $\epee=0$, the $3\sigma $
C.L. contour ends at $(\epet,\eptt)\sim (1.0,1.0)$ for normal
hierarchy, while we have $(\epet,\eptt)\sim (1.8,2.5)$ for inverted
hierarchy. In terms of $\cos^2 \beta$, the contours in Fig.
\ref{atmk2k6panelsnormal} correspond to $\cos^2 \beta \simeq
0.47,0.53,0.65$ (compared with $\cos^2 \beta\simeq 0.30,0.35,0.42$ for
Fig.~\ref{atmk2k6panels}).

From our analytics, we can see two sources of difference between the
results for the two hierarchies.  One traces back to the term
$\lambda_{\tau'}/\Delta $ in the 3-3 entry of the Hamiltonian
(\ref{hamrot}) (see also Eq. (\ref{eff_par})). This can be as large as
$\lambda_{\tau'}/\Delta \sim 0.2$ in the allowed region of parameters
(Eq. (\ref{eq:width})) and so is expected to contribute at the
subdominant level.
%
Secondly, corrections that depend on the sign of $\Delta$ arise also
from the small, but not zero, coupling of the state $\nue^\prime$ with
the other two. This depends on the 1-1 entry of the Hamiltonian
(\ref{hamrot}) and therefore on the relative sign of $\lambda_{e '}$
and $\Delta$.  Considering that the data are dominated by neutrinos
over antineutrinos (due to the larger detection cross section, see
e.g.  \cite{Fukugita:2003en}), it makes sense that a larger allowed
region of NSI is obtained for inverted hierarchy, where, for
neutrinos, matter and vacuum terms have the same sign and thus
suppress the mixing of $\nue^\prime$ more than in the normal hierarchy
case.

\begin{figure}[!b]
\centering
\includegraphics[width=0.47\textwidth]{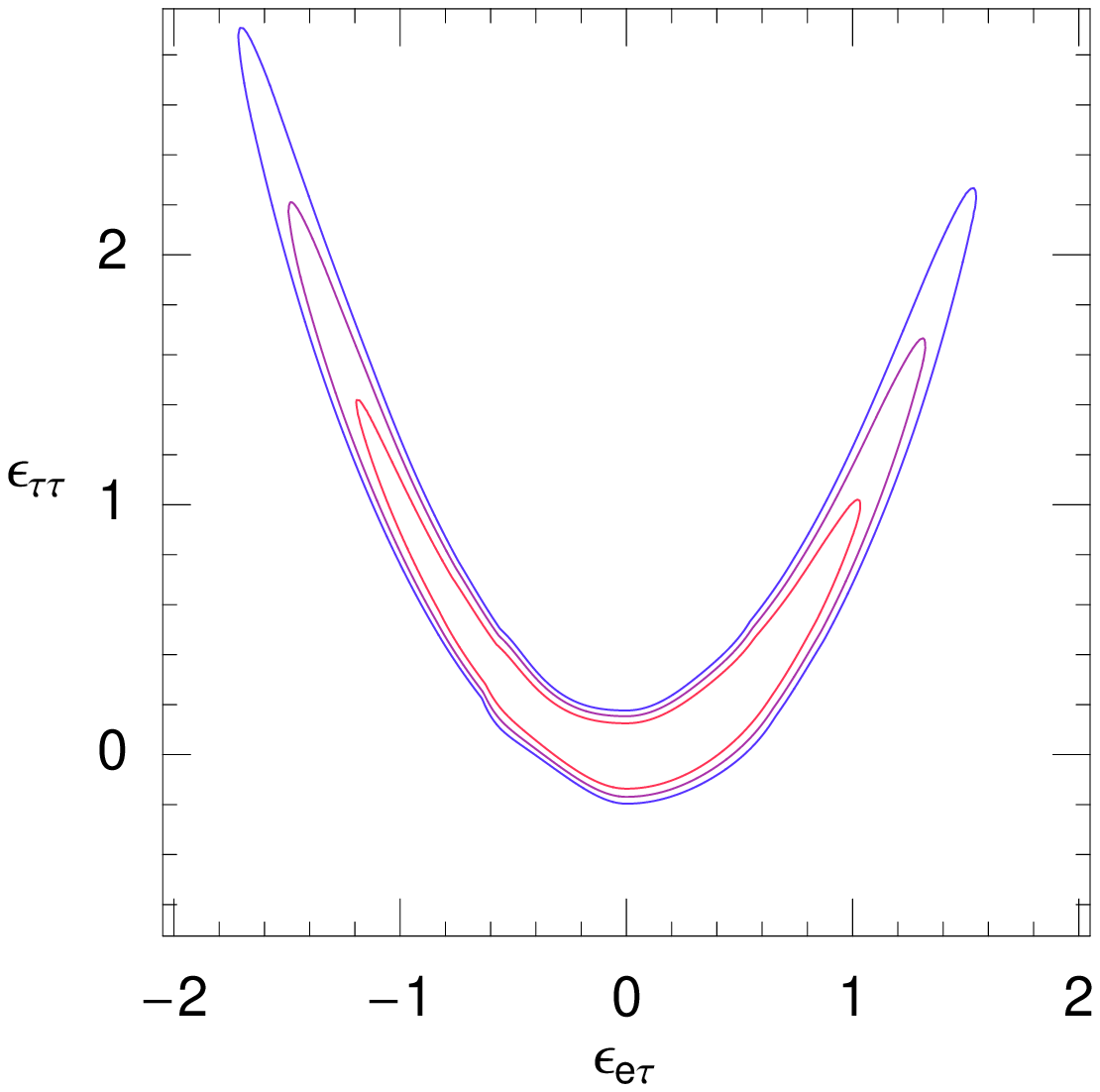}
\caption{An example of how the contours of Fig. \protect\ref{atmk2k6panels} change with non-zero $\theta_{13}$.  This plot refers to $\epee=0$ and $\sin^2 \theta_{13}=0.02$, with $\theta_{13}>0$.}
\label{ee0th13008_k2k}
\end{figure}

Fig.~\ref{ee0th13008_k2k} shows a generalization of Fig.
\ref{atmk2k6panels} to non-zero $\theta_{13}$.  Only the plane
$\epee=0$ is shown for illustration.  The plot was obtained for
$\theta_{13}$ at the reactor limit, Eq. (\ref{th13}), with
$\theta_{13}>0$.  From the comparison of Figs. \ref{ee0th13008_k2k}
and \ref{atmk2k6panels} it is clear that $\theta_{13}$ breaks the
symmetry in the sign of $\epet$.  While the parabolic direction of the
region is unchanged, the position where the region ends along this
direction is affected. These features can be understood analytically,
as shown in Appendix \ref{sect:app}. The effect of non-zero
$\theta_{13}$ on the extent of the allowed region in the space of
$|\Delta m^2_{31}|$-$\sin^2\theta_{23}$ is small and is not shown
here.

Let us now comment on conversion effects in the lowest energy part of
the atmospheric neutrino spectrum, corresponding to the sub-GeV
events. Here oscillations driven by the $\Delta m^2_{21}$ vacuum terms
and matter terms occur on top of faster vacuum oscillations due to
$\Delta m^2_{32}$.  The result is a deviation of the sub-GeV e-like
events from the unoscillated prediction.  This effect has been
discussed in detail for the standard, no-NSI case
\cite{Peres:1999yi,Peres:2003wd,Gonzalez-Garcia:2004cu}.  It was found
that the ratio of the fluxes of neutrinos of muon and electron flavors
depends on $\Delta m^2_{21}$ through the probability $P(\nu_2
\rightarrow \nue)$ of conversion of the mass eigenstate $\nu_2$ into
$\nue$ inside the Earth. This probability is multiplied by the flux
factor ${\cal F}=r \cos^2 \theta_{23}-1 $, with $r$ being the
unoscillated ratio on muon and electron neutrino fluxes.  The
numerical coincidence $\cos^2 \theta_{23} \sim 1/2$ and $r \sim 2$
produces a strong suppression of ${\cal F}$ and thus of the conversion
effect \cite{Peres:1999yi}.

The generalization of this to NSI is immediate.  Given the smallness
of the effect on absolute scale, with an impact on the fit to the data
not larger than few per cents, we choose not to discuss it in detail.
We mention two sources of enhancement of the conversion effect. The
first is a possibly larger flux factor, due to a smaller
$\theta_{23}$, see sec. \ref{mainres}. The second is a larger
probability $P(\nu_2 \rightarrow \nue)$ due to NSI. Indeed, if
flavor-changing NSI are present, $P(\nu_2 \rightarrow \nue)$ converges
to a non-zero value when matter terms dominate over the vacuum ones,
in contrast with the standard scenario (see e.g.
\cite{Friedland:2004pp}).

Finally, it should be pointed out that there exists an important
identification in the parameter space. Indeed, physical results depend
only on the \emph{relative sign} of the vacuum and matter terms of the
Hamiltonian. Because of this, for vanishing $\theta_{13}$, the case of
one hierachy maps onto the case of the other hierarchy with $\epee
\leftrightarrow -2 - \epee $, $\epet \rightarrow -\epet$, $\eptt
\leftrightarrow -\eptt$. This, even though we have presented only the
cases $\epee\ge -1$, our results extend to $\epee< -1$. Explicitly,
normal hierachy with $\epee< -1$ maps to inverted hierarchy with
$\epee> -1$, and, likewise, inverted hierachy with $\epee< -1$ maps to
normal hierarchy with $\epee> -1$. The cases $\epee= -1$ for the two
hierarchies, shown in Figs.~\ref{atmk2k6panels} and
\ref{atmk2k6panelsnormal}, are clearly related by the transformation
$\eptt\leftrightarrow -\eptt$ (and also $\epet\leftrightarrow -\epet$,
but the region is symmetric with respect to this tranformation).

\section{Summary and conclusions} \label{concl}

We have explored the sensitivity of the atmospheric and K2K neutrino
data to neutrino NSI in the $e-\tau$ sector, and investigated how the
presence of NSI can change the allowed region of the vacuum
oscillation parameters.  The results can be summarized as follows:

\begin{enumerate}

\item The region of the NSI parameters $\epee,\eptt,\epet$ allowed by
  the data is essentially determined by two conditions on the
  neutrino-matter interaction Hamiltonian. The first condition is that
  one of the eigenvalues not exceed the vacuum term at high energy
  ($E\sim 20-30$ GeV): $|\lambda_{\tau^\prime}|/\Delta \lta 1 $ or,
  numerically, $|\lambda_{\tau^\prime}|\lta 5 \cdot 10^{-14}$ eV. It
  explains the presence of a ``flat direction'' in the $\chi^2$, which
  extends the allowed region to large NSI couplings. This direction is
  a parabola on planes of constant $\epee$.  Transversely to it, the
  $\chi^2$ function grows rapidly, as can be seen from our figures.
  The second condition is on the mixing angle that describes the
  flavor composition of the eigenstates of the matter Hamiltonian:
  $\beta \lta 0.3 \pi \simeq 57^\circ$. It follows from requiring
  consistency between atmospheric data samples of different energy
  and/or between the atmospheric data and the (practically)
  matter-free K2K results. This condition explains the worsening of
  the fit along the parabolic flat direction.
  
  In terms of the epsilons, we see that the allowed range of $\epet$ and
  $\eptt$ strongly depends on the value of $\epee$, which in turn is
  unconstrained by atmospheric neutrinos. Both $\epet$ and $\eptt$ are
  most constrained for $\epee=-1$, and the constraint rapidly weakens
  as $\epee$ is increased. For $\epee=1.5$ and inverted mass
  hierarchy, values as large as $\eptt \sim 5$ and $|\epet| \sim 3.5$
  are allowed along the parabola. These would translate into very
  loose limits on the NSI parameters of the effective four-fermion
  Lagrangian. Still, such limits generally improve on existing
  accelerator bounds.
%

\item The inclusion of NSI in the analysis modifies the allowed region
  of the vacuum oscillation parameters, $\sin^2 \theta_{23}$ and
  $\Delta m^2_{31}$. As our analytical treatment shows, the fit to
  large NSI is achieved at the expense of changing the values of the
  vacuum oscillation parameters. After marginalizing over the NSI
  couplings and the sign of the mass squared splitting, we find that
  the region in the space of $\sin^2 \theta_{23}$ and $|\Delta
  m^2_{31}|$ is larger than that of the standard, no-NSI case.
  Smaller mixing and larger mass splitting are allowed.  If we fit the
  data to one parameter at a time, and marginalize over all the others
  (see Table \ref{tabmassmix}) we find the intervals $\sin^2
  \theta_{23}=0.24 - 0.68$ and $|\Delta m^2_{31}|=(1.7 - 3.9)\cdot
  10^{-3}~{\rm eV^2}$ at $3\sigma$ C.L. ( to be compared to the
  results of the standard case: $\sin^2 \theta_{23}=0.32 - 0.66$ and
  $|\Delta m^2_{31}|=(1.7 - 3.6)\cdot 10^{-3}~{\rm eV^2}$).

\item The recent K2K results play an important role in limiting NSI.
  This stems from the fact that for the K2K setup matter effects are
  negligible, and therefore K2K measures the true vacuum oscillations
  parameters. The K2K constraint on the oscillation parameters,
  particularly on $|\Delta m^2_{31}|$, translates in a constraint on
  NSI, by limiting the range over which the oscillation parameters
  could be varied to compensate for the effects of large NSI. As an
  example, for $\epee \sim {\cal O}(10^{-1})$, the addition of the K2K
  results restricts the region of the NSI couplings by about $\sim
  25\%$ in the direction of $\eptt$ (Fig. \ref{k2keffectee03}), with
  respect to the analysis of atmospheric neutrinos only.

\item We have studied the dependence of the results on the mass
  hierarchy (sign of $\Delta m^2_{31}$). The allowed region of the NSI
  couplings has the same shape for both hierarchies, but, for $\epee
  \gta -1$, it is more extended for the inverted hierarchy by up to a
  factor of $\sim 2$ in the direction of $\eptt$.  Subdominant effects
  due to $\theta_{13}$ have been analyzed. They mainly break the
  degeneracy in the sign (phase) of $\epet$.  Corrections due to the
  smaller mass squared splitting, $\Delta m^2_{21}$, turn out to be
  quite small.

\end{enumerate}

Our results represent a step toward the reconstruction of the region
in the space of NSI couplings that is compatible with all existing
data.  One of the next steps is the extension of the analysis to
include the solar neutrino and KamLAND data. While it is known that
the latter do not exclude large NSI couplings along the parabolic
region (\ref{parab}) \cite{Friedland:2004pp}, a detailed study has not
been done before; it is presented in a companion paper of this work,
soon to be completed \cite{usprep}.

The fact that still large NSI in the $e-\tau$ sector are not
experimentally excluded has important implications for other aspects
of neutrino physics.  First, it is an important motivation for
neutrino experiments with man-made sources.  Experiments with neutrino
beams of short or intermediate base-line, such as MINOS
\cite{Thomson:2005hm} or OPERA \cite{Autiero:2005hn} will have
negligible matter effects, and will increase the precision of the
measurement of $\theta_{23}$ and $|\Delta m^2_{31}|$. This increased
precision will leave even less room for NSI, or give indication of
their existence, depending on whether the measured vacuum parameters
are in agreement or in tension (especially if the tension is in the
direction of smaller mixing and/or larger mass splitting) with the
analysis of atmospheric neutrino with standard interactions only.

Neutrino beams with energy $E\sim 1-10$ GeV and long base-line (of the
order of thousands of Km) like the proposed Fermilab-to-Soudan design,
for example, would exhibit dramatic effects of NSI in the
disappearance of muon neutrinos (or antineutrinos) as well as in the
appearance channel $\numu \rightarrow \nue$ (or $\barnumu \rightarrow
\barnue$).  In the disappearance channel NSI would produce an
irregular pattern of oscillations minima and maxima, due to all three
neutrinos being involved in the oscillations, in contrast with the
simpler two-neutrino $\numu \rightarrow \nutau$ oscillations expected
in absence of NSI.  The appearance channel $\numu \rightarrow \nue$
would be particularly characteristic in the fact that the $\nue$
component could be much larger than what allowed by standard
interactions and subdominant effects (those of $\theta_{13}$ and of
$\Delta m^2_{21}$) and would not be suppressed at high energy -- in
contrast with the case of the MSW effect with standard interactions --
as a consequence of flavor-changing NSI.  In the high-energy limit,
i.e. where the matter potential in the Earth dominates over vacuum
terms, the amplitude of the $\numu \rightarrow \nue$ oscillation would
be controlled by the matter mixing $\beta$:
\begin{equation}
 P(\numu \rightarrow \nue) \sim \sin^2 \beta
 P(\numu \rightarrow \nu_{\tau}^\prime)~,
 \label{beamprob}
\end{equation}
where $ P(\numu \rightarrow \nutau^\prime)$ is given in Eq.
(\ref{p_hier}) and is unsuppressed at high energy for NSI along the
parabola (\ref{parab}).


The presence of NSI can also alter significantly the physics of
supernova neutrinos, in a way that may be tested with data from a
future galactic supernova. Firstly, in the outer regions of the
collapsing star, the NSI couplings may produce a richer structure of
level-crossings with respect to the two MSW resonances of the standard
case. New resonances may, in principle, appear depending on the
chemical composition of the medium in the star and on what scatterers
are responsible for the NSI. A less trivial effect will be on the
evolution of trapped neutrinos inside the core of the protoneutron
star. Here, NSI may affect the explosion itself, by changing the
dynamics of the core and the energetics of the shock wave. This
fundamental effect has been overlooked until recently
\cite{Amanik:2004vm}. More work is required to understand the full
implications of NSI, including possible changes in the r-processes and
in the energy deposition by neutrinos in the matter of the star. 

\section*{Acknowledgments }
We thank the K2K collaboration (J.~Wilkes and R.~Gran in particular)
for useful information and for sharing with us the results of the K2K
data analysis.  We acknowledge the effort of M. Maltoni in the initial
stage of this work, and thank him for providing the numerical
executable used for our calculations.
A special thank goes to the Oak Ridge National Laboratory for allowing
the use of their numerical resources and to A. Mezzacappa and W.
Haxton for facilitating the contact with ORNL.  A.F. acknowledges
support from the Department of Energy, under contract number
W-7405-ENG-36.  C.L. thanks the IAS of Princeton and LANL for
hospitality during the preparation of this work. She also acknowledges
the INT-SCiDAC grant number DE-FC02-01ER41187 for financial support.

\begin{table*}
\begin{tabular}{|l|l|l|l|l|}
\hline
\hline
C.L. & $\sin^2 \theta_{23}$ & $|\Delta m^2_{31}|/10^{-3}~{\rm eV^2}$ & $\sin^2 \theta_{23}$ (no-NSI) & $|\Delta m^2_{31}|/10^{-3}~{\rm eV^2}$ (no-NSI) \\
\hline
\hline
95\%  & 0.31 - 0.64 & 2.0 - 3.4 & 0.36 - 0.60 &  2.0 - 3.2 \\
\hline
99\%  & 0.26 - 0.66  & 1.8 - 3.7  &  0.34 - 0.64 & 1.8 - 3.5 \\
\hline
$3\sigma$  & 0.24 - 0.68 & 1.7 - 3.9  & 0.32 - 0.66  &  1.7 - 3.6 \\
\hline
\hline
\end{tabular}
\caption{The allowed intervals of the vacuum mixing parameters,
$\sin^2 \theta_{23}$ and $|\Delta m^2_{31}|$. Each interval has been
obtained by marginalizing over all the parameters except the one in
question. The same results for the standard case (no NSI) are shown for comparison.}
\label{tabmassmix}
\end{table*}

\appendix

\section{Estimating corrections due to $\theta_{13}$}
\label{sect:app}

As seen in Fig.~\ref{ee0th13008_k2k}, the effect of nonzero
$\theta_{13}$ is to break the $\epet\leftrightarrow -\epet$ symmetry,
while preserving the general parabolic shape of the region.  These
features can be understood by generalizing our analytical description
of Sec.~\ref{dom}.  The corrections due to $\theta_{13}$ enter the
vacuum part of the Hamiltonian, and therefore influence those
predictions that depend on vacuum terms, like the mixing and mass
splitting in matter, Eq. (\ref{eff_par}), and the ``cutoff''
conditions (\ref{alexcond}) and (\ref{cecicond}).  One expects an
interplay between $\epet$ terms and $\theta_{13}$ terms, since both
couple $\nue$ to $\nutau$.  This interplay has the form of an additive
interference, analogous to the one involving the vacuum terms and the
standard interaction in the MSW effect.  This is the origin of the
breaking of the symmetry in the sign (phase) of $\epet$.

Expanding the Hamiltonian in Eq.~(\ref{hamrot}) along the parabolic
direction (\ref{parab}) to first order in $\sin 2\theta_{13}$, we find
the generalizations of Eqs.  (\ref{vac_mix}) and (\ref{vac_dm2}):
\begin{eqnarray}
&&\cos \theta \simeq   \frac{1}{\sqrt{1+c^2_\beta}}
+ \frac{ \sin 2 \theta_{13} s_{2 \beta} \cos(2\psi-\delta)}{4
  (1+c^2_\beta)},
\nonumber \\
\label{mixbestcorr}
&&\Delta m^2 \simeq \Delta m^2_m \frac{1+\cos^{-2}\beta}{2}
\left[1+\frac{\sin 2\theta_{13} \tan \beta}{(1+\cos^2\beta)^{3/2}}
\right].\;\;\;\;\;\;\;\;
\label{dmbestcorr}
\end{eqnarray}
As in Eqs.  (\ref{vac_mix}) and (\ref{vac_dm2}), these expressions
give the values of the vacuum parameters that correspond to
$\theta_m\simeq \pi/4$ and a given value of the mass splitting in
matter, $\Delta m^2_m$.



The generalized form of the condition (\ref{alexcond}) is
\begin{widetext}
\beq
&&-\sqrt{\frac{1}{\tan^2\theta}-1}-\sin 2\theta_{13} c_\theta \cos(2\psi-\delta) \lta \tan \beta \lta + \sqrt{\frac{1}{\tan^2\theta}-1}-\sin 2\theta_{13} c_\theta \cos(2\psi-\delta) ~.
\label{alex_gen}
\eeq
\end{widetext}
Eqs. (\ref{mixbestcorr})-(\ref{alex_gen}) are accurate to the order
$\sin 2\theta_{13}$, while ${\cal O}(\sin^2 \theta_{13}) $ or higher
terms are not under control.  We have neglected terms proportional to
$\lambda_{\tau'}/\Delta \ll 1$ (which is accurate along the parabola).
The $\theta_{13}$ correction to Eq. (\ref{cecicond}) is more
complicated and will not be given here.  We notice that the
$\theta_{13}$ term in Eqs.  (\ref{mixbestcorr}) comes as a product
with $\sin \beta$ from an expansion of terms of the type $|A \sin
\beta/\Delta + \sin \theta_{13}e^{i \delta}|^2 $; this confirms the
MSW-like interference mentioned above.  One can also see how results
depend on the \emph{relative} phase $2\psi-\delta$ of the matter and
vacuum terms.  For $\theta_{23}$ in the first octant, $\psi=\delta=0$,
and $\theta_{13}>0$ Eq. (\ref{alex_gen}) gives a stronger restriction
for positive rather than negative $\epet$.  Similarly, from
(\ref{dmbestcorr}) we see that a positive $\epet$ would increase the
ratio $\Delta m^2/\Delta m^2_{m}$, thus making the tension with the
low-energy sample and/or with K2K stronger.  Both of these features
correspond to the trend observed in Fig.  \ref{ee0th13008_k2k}.  We
have checked that there is also quantitative agreement between
numerical and analytical results (Eqs.
(\ref{mixbestcorr})-(\ref{alex_gen})) at the order of magnitude level.
%

\end{document}